\documentclass[showpacs,preprintnumbers,amsmath,amssymb,floatfix,prd,11pt]{revtex4}
\pdfoutput = 1
\usepackage[utf8]{inputenc}
\usepackage{epsfig}
\usepackage{amsmath}

\begin{document}
\title{Inclusive-jet and Di-jet Production in Polarized Deep Inelastic Scattering}
\author{Ignacio Borsa}  
\email{iborsa@df.uba.ar}
\affiliation{Departamento de F\'{\i}sica and IFIBA, Facultad de Ciencias Exactas y Naturales, Universidad de Buenos Aires, Ciudad Universitaria, Pabell\'on\ 1 (1428) Buenos Aires, Argentina}
\author{Daniel de Florian}  
\email{deflo@unsam.edu.ar}
\author{Iv\'an Pedron}  
\email{ipedron@unsam.edu.ar}
\affiliation{International Center for Advanced Studies (ICAS), ICIFI and ECyT-UNSAM, 25 de Mayo y Francia, (1650) Buenos Aires, Argentina}

\begin{abstract}

We present the calculation for single-inclusive jet production in (longitudinally) polarized deep-inelastic lepton-nucleon scattering at next-to-next-to leading order (NNLO) accuracy, based on the \textit{Projection-to-Born} method. As a necessary ingredient to achieve the NNLO results, we also introduce the next-to-leading-order (NLO) calculation for the production of di-jets in polarized DIS. Our di-jet calculation is based on an extension of the \textit{dipole subtraction} method to account for polarized initial-state partons. We analyze the phenomenological consequences of higher order QCD corrections for the Electron-Ion Collider kinematics. 

\end{abstract}

\maketitle

\section{Introduction}

Much progress has been made in our understanding of the structure of hadrons over the last decades, both from the theoretical and the experimental sides. The study of the spin structure of hadrons in terms of its components, particularly the proton, is, however, still one of the challenges faced by particle physics. The spin content of the proton can be codified in terms of the polarized parton distributions of quarks and gluons, which can be experimentally probed in high energy collisions processes with polarized nucleons. Contrary to the case of unpolarized parton distributions, which have been extensively studied for a wide kinematical range, based on several complementary measurements from different observables, our knowledge on the helicity distributions for partons inside the proton is more limited. While more than 30 years ago, fixed-target Deep-Inelastic Scattering (DIS) measurements from EMC refuted the naive interpretation of the parton model, proving that the amount of spin carried by quarks and antiquarks is relatively small~\cite{Aidala:2012mv}, the exact decomposition of the proton spin between quarks, gluons and orbital angular momentum is still unclear. Polarized proton-proton collisions performed at the BNL Relativistic Heavy-Ion-Collider (RHIC)~\cite{Aschenauer:2013woa}, which receive significant contributions from gluon-initiated processes, improved the description of the gluon spin distribution, showing that its contribution to the proton spin is not negligible~\cite{deFlorian:2014yva}, although providing constraints only for a reduced range of proton momentum fractions. Furthermore, the amount of spin carried by the  \textit{sea quarks} is also still an open question~\cite{deFlorian:2009vb,Nocera:2014gqa}. In that sense, the future US-based Electron-Ion-Collider (EIC), allowing a much wider kinematical range, and reaching an unprecedented precision for polarized measurements~\cite{Accardi:2012qut}, is expected to provide new insights on the spin decomposition of the proton in terms of its fundamental building blocks~\cite{Aschenauer:2012ve,Aschenauer:2015ata,Aschenauer:2020pdk}.

In addition to high-precision measurements for a wider range of momentum fractions, the improvement of our picture of the proton spin will require a consistent increase in the accuracy of the theoretical description of the observables to be measured. It is known that leading order (LO) perturbative calculations $\mathcal{O}(\alpha_S^0)$ in QCD only provide qualitative descriptions, since higher order corrections in the strong coupling constant are sizable. Although a remarkable effort to compute higher order corrections for unpolarized processes has taken place during the last 30 years, setting next-to-next-to-leading order (NNLO) as the standard for Large-Hadron-Collider (LHC) calculations and even reaching the following order for some processes, the picture for polarized calculations is not as developed.
Polarized calculations in dimensional regularization necessarily involve dealing with extensions of the $\gamma_{5}$ matrix and Levi-Civita tensor to an arbitrary number of dimension, making the computation much more intricate than its unpolarized counterpart. Until recently, NNLO corrections for polarized processes were only obtained for completely inclusive Drell-Yan~\cite{Ravindran:2003gi} and DIS~\cite{Zijlstra:1993sh}, in addition to the helicity splitting functions~\cite{Vogt:2008yw,Moch:2014sna,Moch:2015usa}. More exclusive observables provide results that can be directly compared to experimental data, and could, in principle, be used to disentangle individual contributions associated to different partons. In particular, jet production in DIS is an extremely useful tool to probe the partonic densities, since it can give a stronger grip on the gluon distribution, while avoiding non-perturbative corrections associated to final-state hadronization. Developments in techniques for flavour and charge tagging in jet production could further improve the potential of jet measurements to disentangle individual flavour contributions in global analysis~\cite{Arratia:2020azl,Kang:2020fka}.

Higher order corrections are not only necessary to improve the accuracy of the theoretical description. It is also important to check the stability of the perturbative series, that is, how these corrections affect the resulting cross sections and spin asymmetries, since only processes perturbatively well behaved can be used as good probes for parton distributions, and be utilized for its extraction. Furthermore, for the specific case of jet production, it is only at higher orders in QCD that jet structure is fully developed, allowing to realistically match the theoretical description to the experimental data and the cuts imposed in the jet reconstruction. 

In this paper we present the NLO calculation for di-jet production in polarized and unpolarized lepton-nucleon DIS, based on an extension of the Catani-Seymour dipole subtraction method~\cite{Catani:1996vz} to account for polarized initial particles. We analyze the structure of higher order corrections in the Electron-Ion-Collider kinematics, its perturbative stability and phenomenological implications. Through a detailed study of the polarized cross sections and asymmetries we also identify the most important partonic contributions for different kinematical regions. Additionally, we expand on our previous results~\cite{Borsa:2020ulb} for single-exclusive jet production in DIS at NNLO, achieved by combining the aforementioned di-jet result with the inclusive polarized NNLO DIS structure functions~\cite{Zijlstra:1993sh} through the application of the Projection-to-Born (P2B) method~\cite{Cacciari:2015jma}. We analyze the perturbative stability of the higher order corrections to the cross section and asymmetries, as well as the contributions from the different partons to the NNLO corrections. Both the NLO single- and di-jet, as well as the NNLO single-jet calculations are implemented in our code {\tt{POLDIS}} \footnote{The code is available upon request from the authors}.

The remaining of the paper is organized as follow: in section \ref{sec:kinematics} we begin by defining the kinematics of both single- and di-jet production in DIS. In section \ref{sec:higher_order_corrections} we detail both our extension of the dipole subtraction method for polarized QCD processes, and its use in the P2B method in order to achieve polarized jet production at NNLO. In section \ref{sec:di-jets} we present the phenomenological results for inclusive NLO di-jet production at the EIC in the Breit-frame, and in section \ref{sec:single-jets} we do the same for inclusive NNLO single-jet production in the laboratory frame. Finally, in section \ref{sec:conclusion} we summarize our work and present our conclusions.

\section{Jet production kinematics}\label{sec:kinematics}

We start considering the case of inclusive-single jet production in DIS. Specifically, we study the process $$e(k)+P(p)\rightarrow e(k')+\mathrm{jet}(p_T,\eta)+X, $$ where $k$ and $p$ are the momenta of the incoming electron and proton, respectively, and $k'$ is the momentum of the outgoing electron. We work in the laboratory frame (L), where single-jet production receives non-vanishing contributions already at $\mathcal{O}(\alpha_{S}^{0})$. We only consider, for the time being, neutral-current processes mediated by the exchange of a virtual photon, with its momentum $q=k-k'$ and virtuality $Q^{2}=-q^{2}$ fully determined by the electron kinematics. The inelasticity $y$ and Bjorken variable $x$ are then defined as usual by 

\begin{equation}
    y=\frac{q\cdot p}{k\cdot p}, \qquad x=\frac{Q^{2}}{2 p\cdot q}.
\end{equation}

In addition to the variables commonly used for fully inclusive DIS, more insight on the underlying partonic kinematics can be obtained through the study of the final-state jet, which can be characterized in terms of its transverse momentum $p_{T}$ with respect to the beam, and its pseudorapidity $\eta$.

\begin{figure}[t]
 \epsfig{figure= 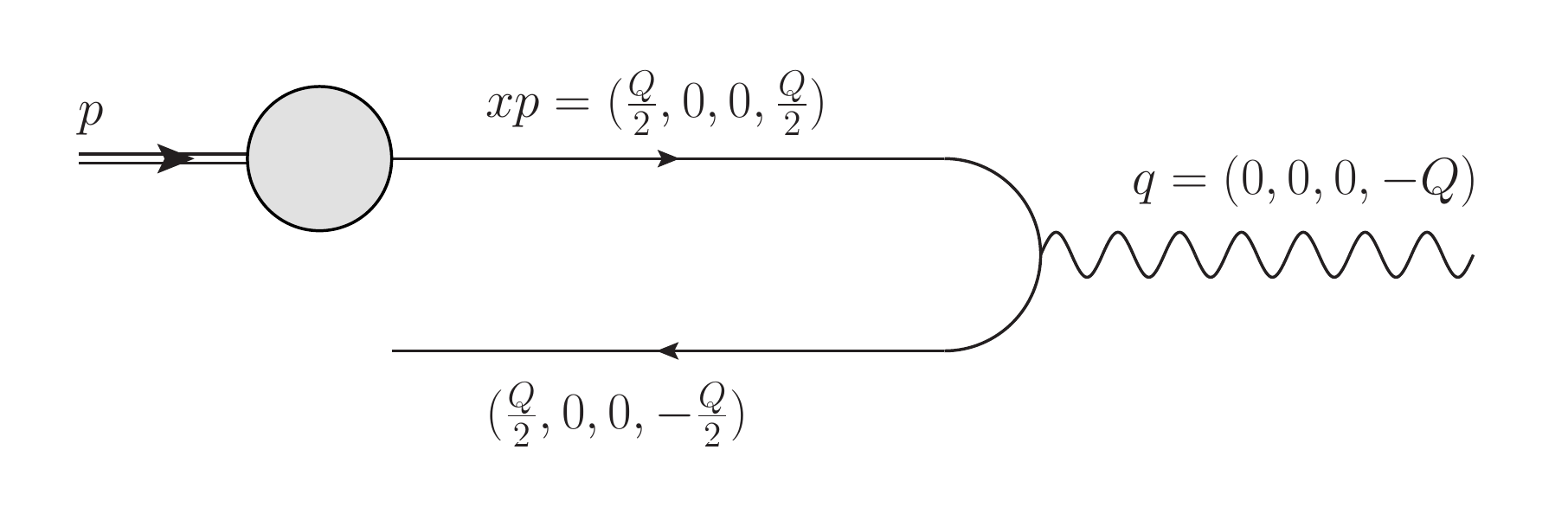 ,width=0.95\textwidth}
  \caption{$\mathcal{O}(\alpha_{S}^{0})$ Breit frame kinematics for the process $p_{1}(xp)+\gamma^{*}(q)\rightarrow p_{2}(xp+q)$.}\label{fig:Breit}
\end{figure}

At higher orders in $\alpha_{S}$, the production of multiple final-state jets becomes available. Di-jet production can be better studied in the Breit frame (B), where there is no contribution of $\mathcal{O}(\alpha_{S}^{0})$ to the production of transverse jets. Formally, the Breit frame is defined as the one that satisfies $2xp+q=0$. Note that for the $\mathcal{O}(\alpha_{S}^{0})$ process, this implies that the virtual photon and incoming quark collide head-on, completely reversing the momentum of the quark (hence the commonly used nickname \textit{brick-wall} frame), as its represented schematically in Fig. \ref{fig:Breit}. The first non-vanishing contribution is then obtained at $\mathcal{O}(\alpha_{S})$, with two final-state partons with opposed transverse momentum.

For di-jet production, we specify the process

$$e(k)+P(p)\rightarrow e(k')+\mathrm{jet}(p_{T,1},\eta_{1})+\mathrm{jet}(p_{T,2},\eta_{2})+X.$$

The availability of a second jet allows for a more in-depth study of the partonic kinematics. As in the H1 \cite{Andreev:2014wwa,Andreev:2016tgi} and ZEUS \cite{Abramowicz:2010cka} experiments, and in addition to the jets transverse momentum and pseudorapidities, the di-jet production cross section can be studied in term of the di-jet variables such as the invariant mass $M_{12}$, the di-jet momentum fraction $\xi_2$, as well as the average momentum $\langle p_{T}\rangle_{2}$ and pseudorapidity difference $\eta^{*}$ in the Breit-Frame, which are defined by

\begin{equation}\label{eq:di-jet_kinematics}
\begin{split}
    M_{12}&=\sqrt{(p_{1}+p_{2})^{2}},\\
    \langle p_{T}\rangle_{2}&=\frac{1}{2}(p_{T,1}^B+p_{T,2}^B),\\
    \eta^{*}&=|\eta_{1}^B-\eta_{2}^B|,\\
    \xi_{2}&=x\big(1+\frac{M^{2}_{12}}{Q^{2}}\big).
\end{split}    
\end{equation}

It is worth noticing that, at the LO of di-jet production, $\xi_{2}$ is the momentum fraction carried by the incoming parton.

\section{Calculation of higher order corrections}\label{sec:higher_order_corrections}

Calculations beyond the leading order in QCD necessarily involves cancellations between the individually divergent pieces coming from infrared real emission and virtual diagrams, in addition to the factorization contributions. In the dimensional regularization scheme the number of dimensions is set to $D=4-2\epsilon$, and those divergences then appear as poles in $\epsilon$. The cancellation between those poles can only be achieved after the integration of each of the divergent parts over its appropriate phase space, thus impeding a direct numerical calculation.

Several methods to numerically compute higher order corrections  were developed over the last three decades. The two main approaches are based on either limiting the phase space integration (\textit{phase space slicing}) in order to avoid the divergent regions, or generating appropriate counter-terms (\textit{subtraction}) to cancel the singularities in each of the pieces of the calculation. For the latter, the proposed counter-term should have the same divergent behaviour as the real and collinear parts, while being simple enough to be integrated analytically in order to cancel the poles coming from the virtual diagrams. 

Many general methods for constructing NLO counterterms were proposed. Among them, the \textit{dipole subtraction} method developed by Catani and Seymour, and based on the dipole factorization formula, allows to calculate any jet production cross section at NLO accuracy. The landscape for the following order is complicated due to the appearance of many more singular configurations, but several methods of varying generality are also available for the computation of NNLO calculations \cite{Catani:2007vq,GehrmannDeRidder:2005cm,Boughezal:2011jf,Cacciari:2015jma,Czakon:2010td,Binoth:2004jv,Anastasiou:2003gr,Somogyi:2006da,Stewart_2010}. In particular, for processes where the Born kinematic can be inferred from external non-QCD particles, the P2B method allows to  obtain N$^{k}$LO differential calculations for a jet observable $\mathcal{O}$, given that the N$^{k}$LO inclusive cross section and the differential N$^{k-1}$LO for $\mathcal{O}+jet$ are known. Consequently, given a NLO DIS di-jet calculation and the NNLO structure functions, we can then  compute the NNLO exclusive single jet cross section.

This exclusive NNLO single-jet calculation is implemented in our code {\tt POLDIS} for both polarized and unpolarized DIS. It allows to compute any infrared safe observable related to single-jet production at NNLO accuracy in the laboratory, as well as to single- and di-jet production in the Breit-frame with NLO precision. The code is partially based on {\tt DISENT}, which implements the Catani-Seymour dipole subtraction method to obtain the NLO single- and di-jet cross sections in unpolarized DIS. Mayor modifications were made in order to include the polarized di-jet computation, using an extended version of the dipole subtraction to account for initial-state polarized particles, as well as the implementation of the P2B subtraction in order to obtain NNLO results. Note that the previously reported bug in {\tt DISENT} in the gluon channel \cite{Antonelli:1999kx,Dasgupta:2002dc, McCance:1999jh,Nagy:2001xb} was fixed along with the modifications (see Appendix \ref{sec_apendixbug}). 

Both the extension of the dipole subtraction as well as the P2B method will be discussed in more detail in sections \ref{sec_subtraction} and \ref{sec_PTB}.

\subsection{The dipole subtraction method for polarized processes }\label{sec_subtraction}

For processes involving (polarized) unpolarized initial-state hadrons, QCD calculations necessarily involve convolutions between partonic cross sections and (helicity) parton distribution functions, (p)PDFs, codifying the (spin) momentum distribution of partons inside that hadron. In the case of DIS scattering, the (polarized) unpolarized hadronic cross sections can be written perturbatively as:

\begin{equation}
(\Delta)\sigma(p) = \sum_{a} \int_{0}^{1} dz\, (\Delta) f_{a}(z,\mu_{F}^{2})\left[(\Delta)\hat{\sigma}_{a}^{LO}(zp)+(\Delta)\hat{\sigma}_{a}^{NLO}(zp,\mu_{F}^{2})+...\right], 
\end{equation}

\noindent where $...$ denotes higher order corrections. The helicity pPDF for a parton $a$ carrying a fraction $z$ of the proton's momentum $p$ is defined as $\Delta f_{a}(z,\mu_{F}^{2})\equiv f^{+}(z)-f^{-}(z)$ , with $f^{+}(z)(f^{-}(z))$ denoting the density of partons of type $a$ and momentum fraction $z$, with their helicities aligned (anti-aligned) with that of the proton. On the other hand, the polarized partonic cross section $\Delta\hat{\sigma}\equiv \frac{1}{2}[\hat{\sigma}^{++}-\hat{\sigma}^{+-}]$ is defined in terms of the difference between the cross sections with the incoming lepton and hadron polarized parallel and antiparallel. Up to NLO, the (polarized) unpolarized $m$-parton cross section is given by 

\begin{equation}
(\Delta)\hat{\sigma}_{a}^{LO}(p)=\int_{m}\,d(\Delta)\hat{\sigma}_{a}^{B}(p),
\end{equation}
\begin{equation}
(\Delta)\hat{\sigma}_{a}^{NLO}(p,\mu_{F}^{2})=\int_{m+1}\,d(\Delta)\hat{\sigma}_{a}^{R}(p)+\int_{m}\,d(\Delta)\hat{\sigma}_{a}^{V}(p)+\int_{m}\,d(\Delta)\hat{\sigma}_{a}^{C}(p),
\label{eq_NLO_hadroninc_xsec}
\end{equation}

\noindent where $d (\Delta) \hat{\sigma}^{B}$ is the (polarized) unpolarized partonic Born cross section, and $d(\Delta)\hat{\sigma}^{R}$ and $d(\Delta)\hat{\sigma}^{V}$ stand for the NLO partonic real-emission and virtual matrix elements, respectively. The last term in  Eq.~(\ref{eq_NLO_hadroninc_xsec}) is associated to the collinear factorization that must be introduced in the case of cross sections involving initial hadrons, to account for the divergences arising from initial-state radiation. 

It is worth noticing that we are working in $D=4-2 \epsilon$ dimensions, and that each of the integrals in Eq.~(\ref{eq_NLO_hadroninc_xsec}) is separately divergent in the limit $\epsilon\rightarrow0$. The calculation of polarized cross sections in dimensional regularization is more involved than its unpolarized counterpart, since the extension of the $\gamma_{5}$ matrix and the Levi-Civita tensor $\epsilon^{\mu\nu\sigma\rho}$ appearing in the helicity projection operators in $D$ dimensions is far from trivial. One way to consistently treat $\gamma_{5}$ and $\epsilon^{\mu\nu\sigma\rho}$ is in the HVBM scheme \cite{THOOFT1972189,Breitenlohner:1977hr}, in which the $D$-dimensional space is separated in the standard four-dimensional subspace, and a $(D-4)$-dimensional subspace. In this scheme, $\epsilon^{\mu\nu\sigma\rho}$ is treated as a genuinely four-dimensional tensor, while $\gamma_{5}$ is such that $\{\gamma_{5},\gamma^{\mu}\}=0$ for $\mu=0,1,2,3$, and $[\gamma_{5},\gamma^{\mu}]=0$ otherwise.

An alternative to numerically compute the partonic cross section in Eq.~(\ref{eq_NLO_hadroninc_xsec}) is the so-called \textit{dipole subtraction method}, introduced by Catani and Seymour \cite{Catani:1996vz} as a general framework for the calculation of NLO jet cross sections. This is the method used to compute the NLO corrections of jet observables in both {\tt DISENT} and {\tt POLDIS}. As in other subtraction-based approaches, the idea behind the procedure is to cancel the infrared singularities that appear in the real, virtual and collinear-factorization pieces of the (polarized) unpolarized cross section, which are integrated in different phase spaces ($m$ particles for the virtual diagrams and $m+1$ for the real-emission diagrams), already at the integral level. That cancellation of divergences is achieved through the introduction of a counterterm $d (\Delta) \sigma^A$ that has the same infrared behaviour (in $D$ dimensions) as $d (\Delta) \sigma^R$. By adding and subtracting this term, the NLO calculation can be rewritten as 

\begin{equation}
\int d (\Delta) \hat{\sigma}^{NLO} = \int_{m+1} \left( d (\Delta) \hat{\sigma}^R - d (\Delta) \hat{\sigma}^A \right) +  \int_{m} \left( d (\Delta) \hat{\sigma}^V + d (\Delta) \hat{\sigma}^C+ \int_1 d (\Delta) \hat{\sigma}^A \right).
\label{eq_dipolemstrfrml}
\end{equation}

In Eq.~(\ref{eq_dipolemstrfrml}) the first integral can be numerically performed in four-dimensions since $d (\Delta) \sigma^A$ acts as a local counter-term of $d (\Delta) \sigma^R$. In the second term the cancellation of $\epsilon$ poles requires the integrals to be performed analytically. 

Clearly, the key of the subtraction method lies in the construction of $d (\Delta) \sigma^A$, which in addition to reproduce the divergent behaviour of $d (\Delta) \sigma^R$ should be simple enough to be analytically integrated. In this case the term is constructed by the use of the dipole factorization formula 

\begin{equation}
d (\Delta) \hat{\sigma}^A = \sum_{dipoles} d (\Delta) \hat{\sigma}^B \otimes d (\Delta) V_{dipole}
\end{equation}

\noindent in the collinear and soft limits, where $\otimes$ stands for the appropriate phase space convolution and sums over color and spin indices. The $(\Delta) V_{dipole}$ are the universal dipole factors that match the infrared singular behaviour of $d (\Delta) \sigma^R$. Note that these terms need to be analytically integrable if $D$-dimensions over the single-parton spaces related to soft and collinear divergences in order to make use Eq.~(\ref{eq_dipolemstrfrml}). The construction of these dipole factors for the unpolarized case was already outlined in detail in Catani and Seymour's paper. We now discuss the extension to the particular case of cross sections with one initial-state polarized parton, required for the calculation of the polarized DIS process. 

Following the same notation introduced by Catani and Seymour, the complete polarized local counterterm $d\Delta\hat{\sigma}_{a}^{A}$ is:

\begin{equation}
\begin{split}
d\Delta\hat{\sigma}_{a}^{A}=&\mathcal{N}_{in}\frac{1}{n_{c}(a)\Phi(p_{a})} \sum_{\{m+1\}}d \phi_{m+1}(p_{1},...,p_{m+1};p_{a})\frac{1}{S_{\{m+1\}}}\\
& \Bigg\{\sum_{\mathrm{pairs}\,i,j}\sum_{k\neq i,j}\Delta\mathcal{D}_{ij,k}\,(p_{1},...,p_{m+1};p_{a})\\
&+\sum_{\mathrm{pairs}\,i,j}\Delta\mathcal{D}_{ij}^{a}\,(p_{1},...,p_{m+1};p_{a})\\
&+\sum_{i}\sum_{k\neq i}\Delta \mathcal{D}^{ai}_{k}\,(p_{1},...,p_{m+1};p_{a})\Bigg\},
\label{eq_pol_counterterm}
\end{split}
\end{equation}

\noindent where the terms $\Delta\mathcal{D}_{ij,k}$, $\Delta\mathcal{D}_{ij}^{a}$ and $\Delta \mathcal{D}^{ai}_{k}$ represent the dipole subtraction terms for final-state singularities with a final-state spectator, final-state singularities with an initial-state spectator, and initial-state singularities, respectively. The sum is performed over all the possible $m+1$ final-state partons configurations, with $d\phi_{m+1}$ denoting the corresponding phase space. Additionally, the $1/n_c$ accounts for the average over the initial-state colors, $\Phi(p_{ a})$ is flux factor, and $S_{\{m\}}$ is the Bose symmetry factor for identical particles in the final-state. In $\mathcal{N}_{in}$ the rest of the QCD-independent factors are included.

It is important to note that to create local counter-terms for the polarized DIS NLO cross section, only the polarization of the initial-state parton needs to be considered. In this case, instead of taking the average of its polarizations, the difference between them is used. Spin states of final-state parton are summed over and therefore they are treated as unpolarized. Thus, the dipole subtraction terms $\Delta\mathcal{D}_{ij,k}$ and $\Delta\mathcal{D}_{ij}^{a}$, associated to final-state singularities, are constructed as in ref~\cite{Catani:1996vz} (using the corresponding polarized Born cross section). New expressions for the dipole formulas are therefore only needed in the case of initial-state singularities with one initial-state parton, represented by $\Delta\mathcal{D}^{ai}_{k}$.

As in the case of the unpolarized cross sections, the terms $\Delta\mathcal{D}^{ai}_{k}$ can be obtained from the dipole factorization formula. In the limit $p_a \cdot p_i \rightarrow 0$, where $p_a$ is the moment of the initial-state parton and $p_i$ a final-state one, the dipole factorization formula for the polarized $(m+1)$-parton matrix element can be expressed as

\begin{equation}
_{m+1,a}\langle 1,...,m+1;\Delta a||1,...,m+1; \Delta a \rangle_{m+1,a} = \sum_{k \neq i} \Delta \mathcal{D}^{ai}_k \left( p_1,...,p_{m+1};p_a\right)+...,
\end{equation}

\noindent where $|1,...,m;\Delta a\rangle_{m,a}$ represents an $m$-particle state in the color and helicity space, with $\Delta a$ denoting that the difference between the incoming parton polarizations is considered. The $\sum_k \Delta \mathcal{D}^{ai}_k$ stands for the sum of the polarized dipole contributions, in which the partons $a$ and $i$ act as a single initial-state parton $\widetilde{ai}$, the `emitter', and the final-state parton $k$ acts as the `spectator' $\tilde{k}$. The ... stands for the other non-singular terms in the $p_a \cdot p_i \rightarrow 0$ limit. Each dipole contribution is given by

\begin{equation}
\begin{split}
\Delta \mathcal{D}^{ai}_k \left( p_1,...,p_{m+1};p_a\right) =& -\frac{1}{2 p_a \cdot p_i} \frac{1}{x_{ik,a}}\\
&\cdot_{m,a}\langle 1,...,\tilde{k},...,m+1;\Delta \widetilde{ai}| \frac{\boldsymbol{T}_k \cdot \boldsymbol{T}_{ai}}{\boldsymbol{T}_{ai}^2} \Delta \boldsymbol{V}^{ai}_k |1,...,\tilde{k},...,m+1;\Delta \widetilde{ai} \rangle_{m,a}.
\label{eq_Daik}
\end{split}
\end{equation}

\noindent The $\boldsymbol{T}$ are the color charge operators corresponding to each parton. The emitter and spectator momenta are given respectively by $\widetilde{p}_{ai}^{\mu}=x_{ik,a} p_a^{\mu}$ and $\tilde{p}_{k}^{\mu}=p_k^{\mu}+p_i^{\mu}-(1-x_{ik,a}) p_a^{\mu}$, where

\begin{equation}
x_{ik,a} = \frac{p_k p_a+p_i p_a-p_i p_k}{p_k p_a+p_i p_a}.
\end{equation}

The splitting functions $\Delta \boldsymbol{V}^{ai}_k$ are the only new blocks needed for the extension of the dipole subtraction formalism to the polarized case. They are constructed so that they give the correct eikonal factors in the soft limits, and the correct $D$-dimensional polarized Altarelli-Parisi splitting functions $\Delta P_{ij}$ in the corresponding collinear limits. Similarly to $\Delta P_{ij}$, $\Delta \boldsymbol{V}^{ai}_k$ are matrices in the helicity space of the emitter parton $\widetilde{ai}$, and their expression in given by:

\begin{equation}
\langle s | \Delta \boldsymbol{V}^{q_a g_i}_k | s' \rangle = 8 \pi \mu^{2 \epsilon} \alpha_s C_F \left[\frac{2}{1-x_{ik,a}+u_i}-(1+x_{ik,a})+3 \epsilon (1-x_{ik,a}) \right] \delta_{s s'},
\end{equation}

\begin{equation}
\langle s | \Delta \boldsymbol{V}^{g_a \overline{q}_i}_k | s' \rangle = 8 \pi \mu^{2 \epsilon} \alpha_s T_R \left[2 x_{ik,a}-1-2 \epsilon (1-x_{ik,a}) \right] \delta_{s s'},
\end{equation}

\begin{equation}
\langle \mu | \Delta \boldsymbol{V}^{q_a q_i}_k | \nu \rangle = 8 \pi \mu^{2 \epsilon} \alpha_s C_F \left\{ i \frac{(1-u_i)}{p_a p_k} \left[ x_{ik,a}+2(1-x_{ik,a})(1+\epsilon)\right]  \epsilon^{\alpha \beta \mu \nu} \left(p_i^{\alpha}+p_k^{\alpha} \right) p_a^{\beta} \right\} ,
\end{equation}

\begin{equation}
\begin{split}
\langle \mu | \Delta \boldsymbol{V}^{g_a g_i}_k | \nu \rangle &= 16 \pi \mu^{2 \epsilon} \alpha_s C_A \\ &\left\{i \frac{(1-u_i)}{p_a p_k} \left[ \frac{1}{1-x_{ik,a}+u_i}-1+2(1-x_{ik,a})(1+\epsilon) \right]  \epsilon^{\alpha \beta \mu \nu} \left(p_i^{\alpha}+p_k^{\alpha} \right) p_a^{\beta} \right\},
\end{split}
\end{equation}

\noindent where $u_i= p_a p_i/(p_a p_i+p_a p_k)$. 

Notice, however, that these expressions of the splitting functions $\Delta \boldsymbol{V}^{ai}_k$ as matrices in the helicity states of $\widetilde{ai}$ are not really needed in the polarized case since the spin structure is trivial for both quarks and gluons. This is due to the fact that the spin correlation terms cancel out due to parity conservation in polarized processes (See Appendix \ref{sec_apendixcorr}). Therefore, only the difference between the possible spin states of the emitter parton $\widetilde{ai}$ are required to perform the subtraction. In the case of a gluon emitter, this accounts for the contraction with the tensor $i\, \epsilon^{\rho \sigma \mu \nu}\,p_{a}^{\rho}\,n^{\sigma}/(2 \ p_{a}\cdot n)$, where $n$ is any light-like vector that satisfies $n \cdot p_{a} \neq 0$, while for a quark emitter the tensor $\delta ^{s\,s'}/2$ is used. The resulting kernels are:

\begin{equation}
\frac{\Delta \boldsymbol{V}^{q_a g_i}_k (x;u)}{8 \pi \alpha_s \mu^{2 \epsilon}} = C_F \left[\frac{2}{1-x+u}-(1+x)+3 \epsilon (1-x) \right],
\end{equation}

\begin{equation}
\frac{\Delta \boldsymbol{V}^{g_a \overline{q}_i}_k (x)}{8 \pi \alpha_s \mu^{2 \epsilon}} = T_R \left[2 x-1-2 \epsilon (1-x) \right],
\end{equation}

\begin{equation}
\frac{\Delta \boldsymbol{V}^{q_a q_i}_k (x)}{8 \pi \alpha_s \mu^{2 \epsilon}} = C_F  \left[ x+2(1-x)(1+\epsilon)\right] ,
\end{equation}

\begin{equation}
\frac{\Delta \boldsymbol{V}^{g_a g_i}_k (x;u)}{8 \pi \alpha_s \mu^{2 \epsilon}} = 2 C_A \left[ \frac{1}{1-x+u}-1+2(1-x)(1+\epsilon) \right].
\end{equation}

In order to integrate the dipole subtraction term, $\int_{m+1} d \sigma^A$, the $D$-dimensional integrals of the $\Delta \boldsymbol{V}^{ai}_k$ terms over the dipole phase space $dp_i(\tilde{p_k};p_a,x)$ are needed. The procedure to obtain them is the same one outlined by Catani and Seymour. The resulting expressions $\Delta \mathcal{V}^{a,ai}$ are: 

\begin{equation}
 \Delta \mathcal{V}^{q g} (x; \epsilon)  = -\frac{1}{\epsilon} \Delta P^{g q} (x,0)+ \Delta P^{g q} (x,0) \ln (1-x) - 2 C_F (1-x) + \mathcal{O}(\epsilon),
\end{equation}

\begin{equation}
 \Delta \mathcal{V}^{g q} (x; \epsilon)  = -\frac{1}{\epsilon} \Delta P^{q g} (x,0)+ \Delta P^{q g} (x,0) \ln (1-x) + 2 T_R (1-x) + \mathcal{O}(\epsilon),
\end{equation}

\begin{equation}
\begin{split}
 \Delta \mathcal{V}^{q q} (x; \epsilon)  = &-\frac{1}{\epsilon} \Delta P^{q q} (x,0)+ \delta(1-x)\,C_{F} \left[\frac{1}{\epsilon^{2}}+\frac{3}{2\epsilon}+\frac{\pi^{2}}{6} \right] \\
 &+ C_{F}\Big[-\Big(\frac{4}{1-x} \ln\frac{1}{1-x}\Big)_{+}-\frac{2}{1-x}\ln(2-x)\\ & -3(1-x)-(1+x)\ln(1-x)\Big]+ \mathcal{O}(\epsilon),
\end{split}
\end{equation}

\begin{equation}
\begin{split}
 \Delta \mathcal{V}^{g g} (x; \epsilon)  = &-\frac{1}{\epsilon} \Delta P^{g g} (x,0)+ \delta(1-x)\Big[C_{A}\Big(\frac{1}{\epsilon^{2}}+\frac{11}{6\epsilon}+\frac{\pi^{2}}{6}\Big)+N_{f}T_{R}\Big(-\frac{2}{3 \epsilon}\Big)\Big] \\
 &+ C_{A}\Big\{-\Big(\frac{4}{1-x}\ln\frac{1}{1-x}\Big)_{+}-\frac{2}{1-x}\ln(2-x)-4(1-x)\\
 &+2\Big[-1+x(1-x)+\frac{1-x}{x}-\frac{(1-x)^3}{x}\Big]\ln(1-x)\Big\}+ \mathcal{O}(\epsilon),
\end{split}
\end{equation}

\noindent where $x$ is the phase space convolution variable and the $\Delta P^{a b} (x,0)$ are the aforementioned polarized four-dimensional Altarelli-Parisi kernels. In the HVBM scheme they are given by \cite{Vogelsang:1996im}:  

\begin{equation}
 \Delta P^{q q} (x,\epsilon)  = C_{F} \left[ \frac{2}{(1-x)_+}-(1+x)+3\epsilon(1-x)+\frac{3+\epsilon}{2} \delta(1-x) \right],
\end{equation}

\begin{equation}
 \Delta P^{q g} (x,\epsilon)  = T_R \left[2 x-1-2 \epsilon (1-x) \right], 
\end{equation}

\begin{equation}
 \Delta P^{g q} (x,\epsilon)  = C_{F} \left[2-x+2 \epsilon (1-x) \right],
\end{equation}

\begin{equation}
 \Delta P^{g g} (x,\epsilon)  = 2 C_{A} \left[ \frac{1}{(1-x)_+}-2 x+1+2 \epsilon (1-x)\right]+\left(\frac{\beta_0}{2}+\frac{C_A}{6} \epsilon \right) \delta(1-x).
\end{equation}

A final remark must be made about the polarized factorization counterterms $d\Delta\hat{\sigma}_{a}^{C}$ in Eq.~(\ref{eq_NLO_hadroninc_xsec}). These counterterms are explicitly written as:

\begin{equation}
d\Delta\hat{\sigma}_{a}^{C}=-\frac{\alpha_{S}}{2\pi}\frac{1}{\Gamma(1-\epsilon)}\sum_{b}\int_{0}^{1}dz\big[-\frac{1}{\epsilon}\big(\frac{4\pi\mu^{2}}{\mu_{F}^{2}}\big)^{\epsilon} \Delta P^{ab}(z,\epsilon)+\Delta K^{ab}_{F.S}\big]\,d\sigma_{b}^{B}(zp),
\label{col_counterterm}
\end{equation}

\noindent where the value of $\Delta K^{ab}_{F.S}$ determines the factorization scheme. We work in the conventional polarized $\overline{\mathrm{MS}}$ factorization scheme in which one needs to compensate for the difference between the polarized and unpolarized quark splitting functions  ($\Delta P^{q q} (x,\epsilon)$ and $P^{q q} (x,\epsilon)$, respectively) in $D-$dimensions.
 Since the difference between the two kernels is given by $\Delta P^{q q} (x,\epsilon)-P^{q q} (x,\epsilon)=4 \epsilon C_F (1-x)$, this is equivalent to setting $\Delta K^{qq}_{F.S}=4 C_F (1-x)$ and $\Delta K^{ab}_{F.S}=0$ otherwise in Eq.~(\ref{col_counterterm}).

\subsection{The Projection-to-Born method}\label{sec_PTB}

As it was mentioned, the P2B method allows to obtain the $\text{N}^{k}$LO calculation for a differential observable, provided that its inclusive cross section at that order, as well as the differential cross section for the observable plus a jet are known at $\text{N}^{k-1}$LO. The idea behind the method is to cancel the most divergent parts by simply using the full matrix element at each phase space point as a counterterm, but binning it in an equivalent Born-projected kinematics of the leading order process (hence the name ``Projection-to-Born''). That is, for each event with weight $w$, a counterterm with weight $-w$ is generated, but with the measurement function evaluated in the kinematics of an equivalent leading order process. Note that this requires the Born kinematics to be fully determined by external non-QCD particles. 

The differential cross section for an observable $\mathcal{O}$ at N$^{k}$LO accuracy can be written as:

\begin{equation}\label{eq:proj_to_Born}
    d\sigma^{N^{k}LO}_{\mathcal{O}}= d\sigma^{N^{k-1}LO}_{\mathcal{O}+\mathrm{jet}}-d\sigma^{N^{k-1}LO}_{\mathcal{O}_{\mathrm{P2B}}+\mathrm{jet}}+d\sigma^{N^{k}LO}_{\mathcal{O} \mathrm{incl}},
\end{equation}

\noindent where in $d\sigma^{N^{k-1}LO}_{\mathcal{O}+jet}$ infrared cancellation at the $\text{N}^{k-1}$LO level has already taken place (numerical implementations beyond leading-order thus require the use of an additional subtraction method). It should be noted that as the final-state partons approach the most singular regions, they are clustered in a jet configuration with Born kinematics, and thus the born-projected counterterms exactly cancel the divergent behaviour of the cross section. The appearance of the inclusive term in Eq.~(\ref{eq:proj_to_Born}) is due to the fact that the sum of all the projected events in a given Born phase space point is equivalent to the integration of the additional final-state partons associated with real emission. Thus, the combination of the Born-projected terms along with the $k$-loop virtual diagrams results in the full contribution of the inclusive $\text{N}^{k}$LO cross section to the observable under consideration.

Clearly, the key of the P2B method lies in the kinematical mapping $\mathcal{O} \rightarrow \mathcal{O}_{P2B}$. In the Born level DIS kinematics the momenta of the incoming and outgoing partons are fully determined by the lepton kinematics. The incoming parton has momenta $p=x P$, and the outgoing one $p'=x P+q$. So the mapping to the Born kinematics is given by using these parton momenta to evaluate the measurement function for the born-projected counterterms. Note that this mapping only works in jet production in the laboratory frame, since in the Breit-frame the first non-vanishing contributions starts at order $\mathcal{O}(\alpha_s)$, with two final-state partons (and hence no mapping is possible in terms of $P$, $x$, and $q$).  

In the particular case of single jet production in unpolarized (polarized) DIS at NNLO, the corresponding counterterms are generated from the double-real and one-loop real radiation matrix elements. The combination of those counterterms with the two-loop matrix elements is then equal to the unpolarized (polarized) DIS inclusive cross section at NNLO \cite{Vermaseren:2005qc,Zijlstra:1993sh,Zijlstra:1992qd}. As mentioned, a numerical implementation of the calculation has yet to deal with the sub-leading divergences coming from the single-real radiation and one-loop diagrams contributing to the unpolarized (polarized) di-jet cross section at NLO. Those missing blocks can then be calculated with the implementation of the Catani-Seymour dipole formalism, whose extension to the polarized case was discussed in \ref{sec_subtraction}. We can then re-write Eq.~(\ref{eq:proj_to_Born}) for the production of jets in unpolarized (polarized) DIS at NNLO in terms of the counterterms of Eq.~(\ref{eq_dipolemstrfrml}) as: 

\begin{equation}
\begin{split}
 (\Delta)\hat{\sigma}^{NNLO}_{\mathcal{O}} &=(\Delta)\hat{\sigma}^{NLO}_{\mathcal{O}+jet}-(\Delta)\hat{\sigma}^{NLO}_{\mathcal{O}_{P2B}+jet}+(\Delta)\hat{\sigma}^{NNLO}_{\mathcal{O} \mathrm{incl}}\\
 &=\int_{m+1} \left[ d(\Delta) \hat{\sigma}_{\mathcal{O}+\mathrm{jet}}^R - d (\Delta)\hat{\sigma}_{\mathcal{O}+\mathrm{jet}}^A \right]\\&
 +  \int_{m} \left[ d (\Delta)\hat{\sigma}_{\mathcal{O}+\mathrm{jet}}^V+ d(\Delta)\hat{\sigma}_{\mathcal{O}+\mathrm{jet}}^C + \int_1 d(\Delta) \hat{\sigma}_{\mathcal{O}+\mathrm{jet}}^A \right]\\
& -\int_{m+1} \left[ d(\Delta) \hat{\sigma}_{\mathcal{O}_{\mathrm{P2B}}+\mathrm{jet}}^R - d (\Delta)\hat{\sigma}_{\mathcal{O}_{\mathrm{P2B}}+\mathrm{jet}}^A \right]\\&
-  \int_{m} \left[ d (\Delta)\hat{\sigma}_{\mathcal{O}_{\mathrm{P2B}}+\mathrm{jet}}^V+ d(\Delta)\hat{\sigma}_{\mathcal{O}_{\mathrm{P2B}}+\mathrm{jet}}^C + \int_1 d(\Delta) \hat{\sigma}_{\mathcal{O}_{\mathrm{P2B}}+\mathrm{jet}}^A \right]\\
&+(\Delta)\hat{\sigma}^{NNLO}_{\mathcal{O} \mathrm{incl}},
\label{eq:proj_to_born_completa}
\end{split}
\end{equation}

\noindent where we have used that the inclusive part can be expressed in terms of the P2B counterterms and the double-virtual matrix element for the observable $\mathcal{O}$ as: 

\begin{equation}
    (\Delta)\hat{\sigma}^{NNLO}_{\mathcal{O} \mathrm{incl}}=\int_{m+1} d(\Delta) \hat{\sigma}_{\mathcal{O}_{\mathrm{P2B}}+\mathrm{jet}}^R +\int_{m} \left[ d (\Delta)\hat{\sigma}_{\mathcal{O}_{\mathrm{P2B}}+\mathrm{jet}}^V+ d(\Delta)\hat{\sigma}_{\mathcal{O}_{\mathrm{P2B}}+\mathrm{jet}}^C\right]+
    \int_{m-1} d (\Delta)\hat{\sigma}_{\mathcal{O}}^{VV}.
\end{equation}

\noindent In addition, the complete expression for the counterterm $d (\Delta)\hat{\sigma}_{\mathcal{O}+\mathrm{jet}}^A$ is that given by Eq.~(\ref{eq_pol_counterterm}).

\section{Results of Polarized NLO Di-jet Production}\label{sec:di-jets}

The first step to reach NNLO accuracy for jet production in DIS lies in the calculation of the NLO di-jet cross section. Precisely, in this section we present our results for polarized inclusive di-jet production at NLO in the Breit frame (B). We consider the Electron-Ion Collider kinematics, with beam energies of $E_{e}=18$ GeV and $E_{p}=275$ GeV, and reconstruct the jets with the anti-$k_{T}$ algorithm and $E$-scheme recombination ($R=1$). Furthermore, for di-jet production we fix the normalization and factorization scales central values as $\mu_{F}^{2}=\mu_{R}^{2}=\frac{1}{2}(Q^{2}+\langle p^{B}_{T}\rangle^{2}_{2})\equiv\mu_{0}^2$, with $\alpha_s$ evaluated at NLO accuracy with $\alpha_s(M_z)=0.118$, and require that the pair of leading jets satisfy the kinematical cuts:

\begin{equation}
\begin{tabular}{ c }
    $p^{B}_{T,1}>5\ \mathrm{GeV}$,\\
    $p^{B}_{T,2}>4\ \mathrm{GeV}$,\\
    $|\eta^{L}|<3.5$,
\end{tabular}
\end{equation}

\noindent with the $\eta$ cut imposed in the laboratory frame, while the lepton kinematics is restricted by

\begin{equation}
\begin{tabular}{ c }
    $0.2<y<0.6$,\\
    $25\, \mathrm{GeV}^{2}<Q^{2}<2500 \, \mathrm{GeV}^{2} $.
\end{tabular}
\end{equation}

\begin{figure}
 \epsfig{figure= 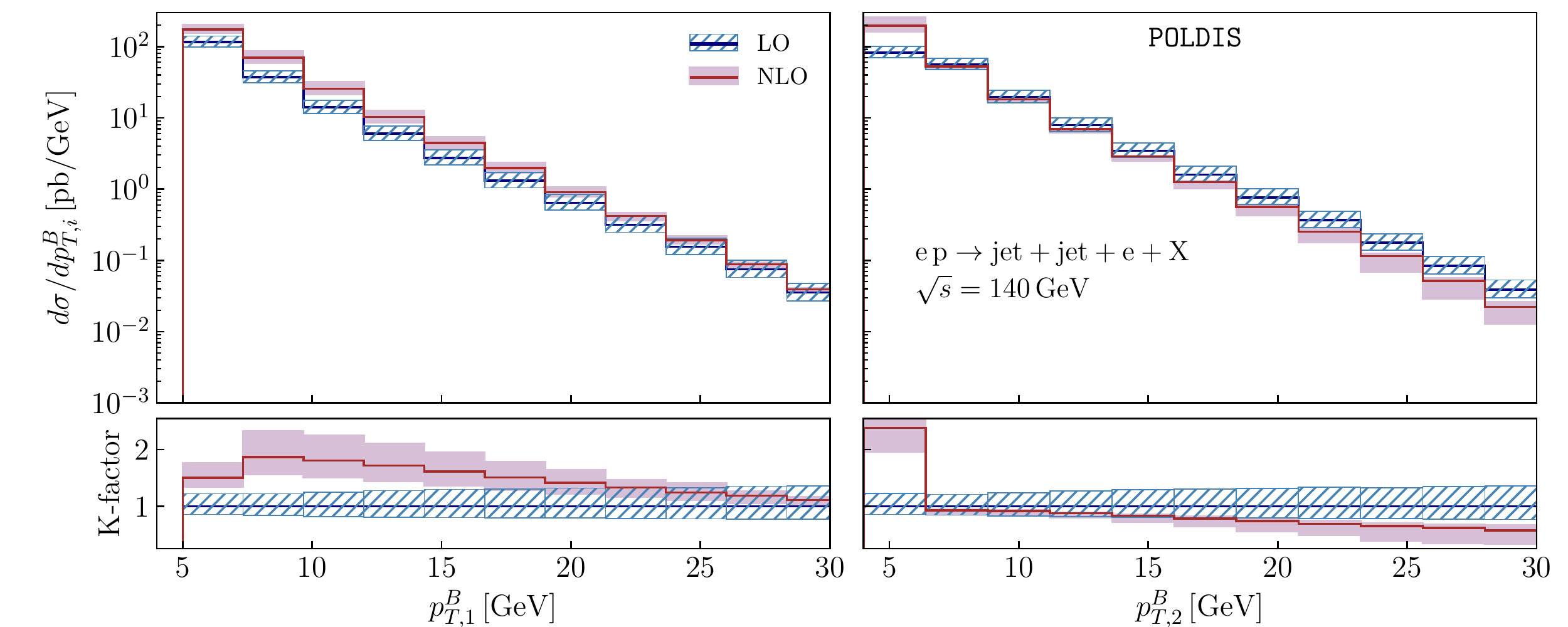,width=0.95\textwidth}
 \epsfig{figure= 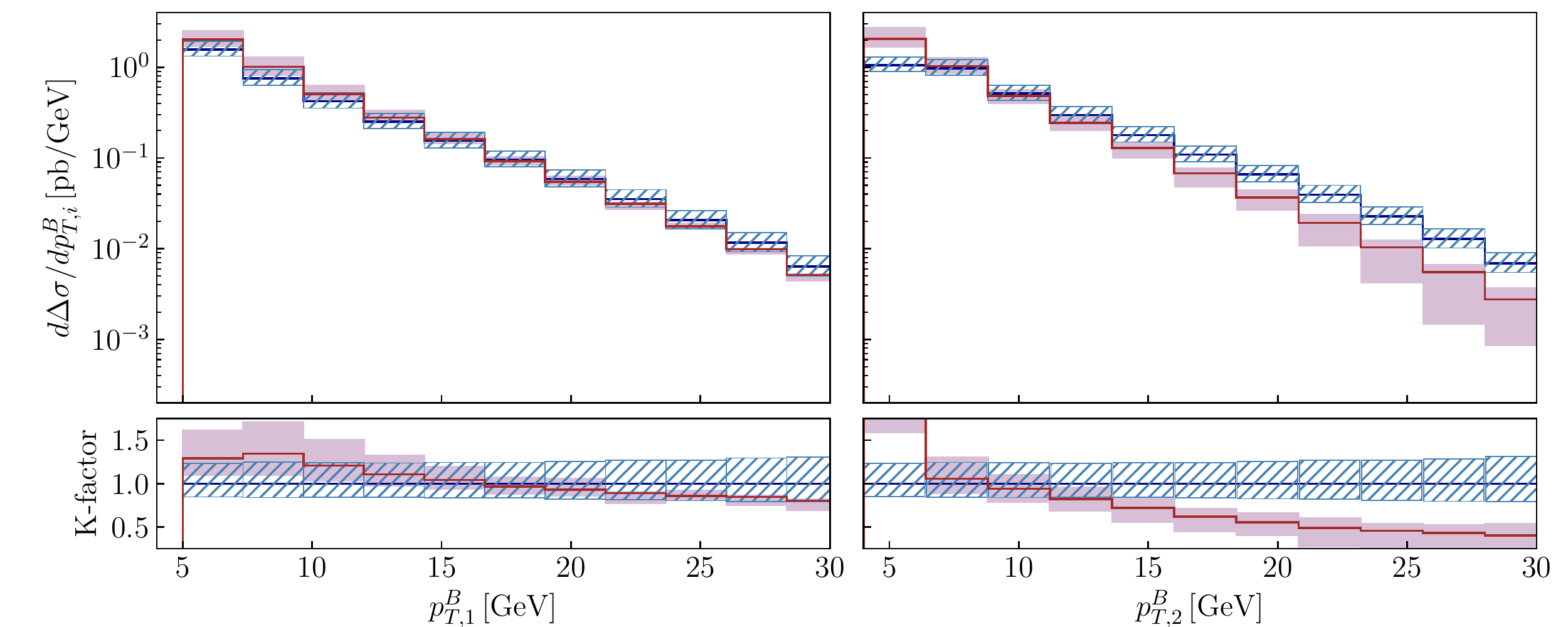,width=0.95\textwidth}
 \caption{Inclusive di-jet production distributions as a function of the leading and sub-leading jet transverse momentum, $p^B_{T,1}$ and $p^B_{T,2}$, respectively, for both the polarized and unpolarized cases. The bands reflect the seven point variation in the cross section when independently changing the scales as $\mu_R,\mu_F= [1/2,2]\frac{1}{2}\mu_{0}$. The lower inset for each plot depicts the K-factor, defined as ther ratio to the LO cross section.}\label{fig_dist_pt12_pol}
\end{figure}

The parton distributions sets used were the NLOPDF4LHC15 \cite{Butterworth:2015oua} and DSSV \cite{deFlorian:2014yva,deFlorian:2019zkl} for the unpolarized and polarized case, respectively.

We begin by presenting the LO and NLO results for the unpolarized and polarized cross sections in Fig. \ref{fig_dist_pt12_pol}, in terms of the leading and sub-leading jet transverse momentum, $p_{T,1}^B$ and $p_{T,2}^B$, respectively. The lower inset in each plot shows the corresponding $K$-factor, defined as the ratio to the LO cross section $\sigma_{LO}$, in order to quantify the effect of the NLO corrections. The bands presented in Fig. \ref{fig_dist_pt12_pol} represent the estimation for the theoretical uncertainties, obtained by independently varying the renormalization and factorization scales as $\mu_R,\mu_F = [1/2,2]\, \mu_{0}$ (with the additional constrain $1/2\leq\mu_{F}/\mu_{R}\leq2$). 

For all the distributions, rather large NLO corrections are obtained, particularly for the low $p^B_{T}$ bins. These sizable corrections are associated to the asymmetric cuts chosen for the $p^B_{T}$ of the leading and sub-leading jets, which can already be noted in the different coverage of each of the distributions. It is also worth noticing the difference in sign of the NLO corrections, which enhance the distributions of the leading jet, while suppressing the sub-leading jet distributions. Similar comments can be made for the polarized distributions in Fig. \ref{fig_dist_pt12_pol}. Compared with the unpolarized case, they show a milder enhancement of the leading jet distribution and a stronger suppression of the sub-leading jet distribution. 

\begin{figure}
 \epsfig{figure= 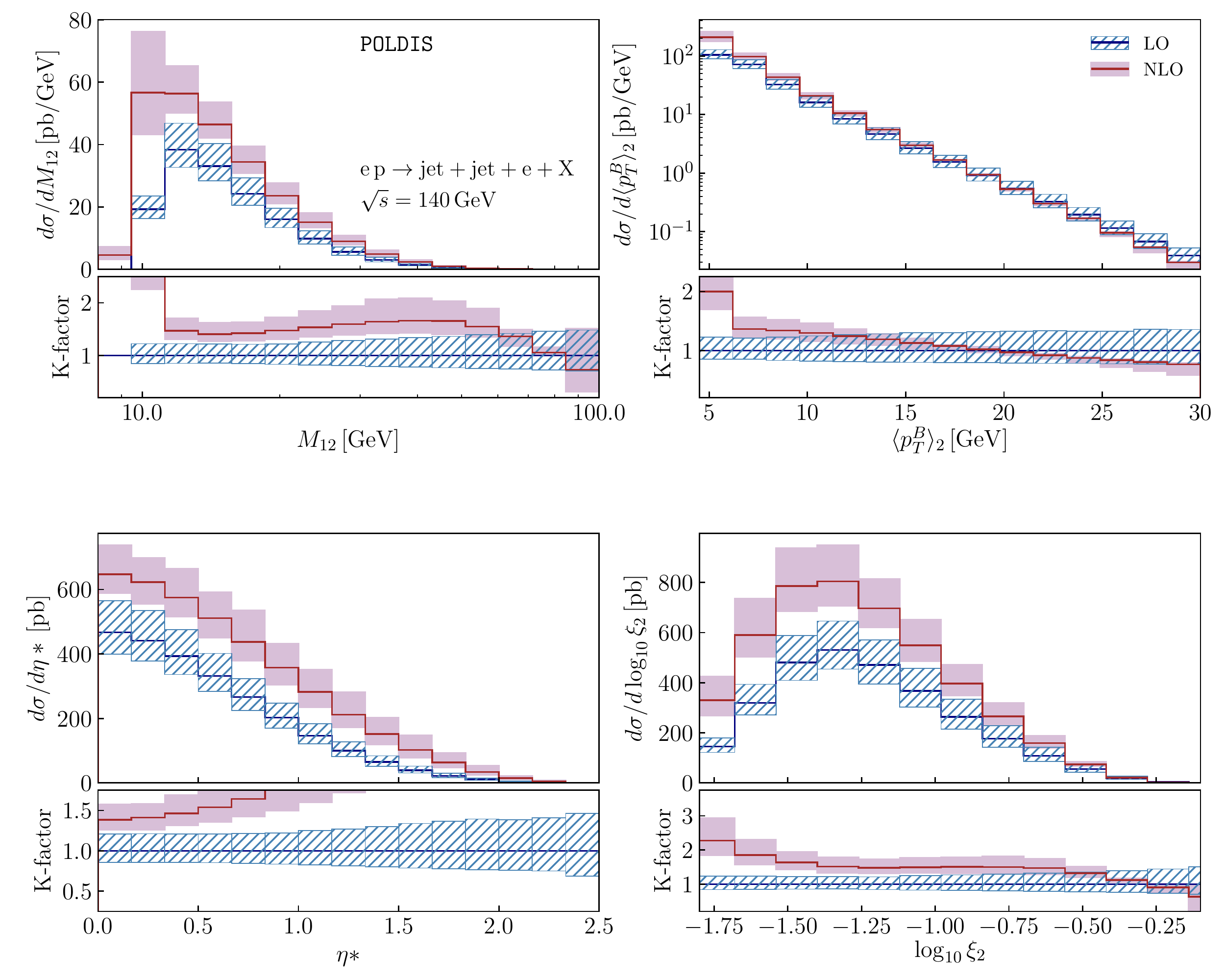, width=0.9\textwidth}
  \caption{Inclusive di-jet production distributions as a function of the variables $\langle p_T^B \rangle_2$, $M_{12}$, $\eta^*$ and $\log_{10}(\xi_2)$. The lower boxes show the K-factor for each distribution.}\label{fig_dist_nopol}
\end{figure}

\begin{figure}
 \epsfig{figure= 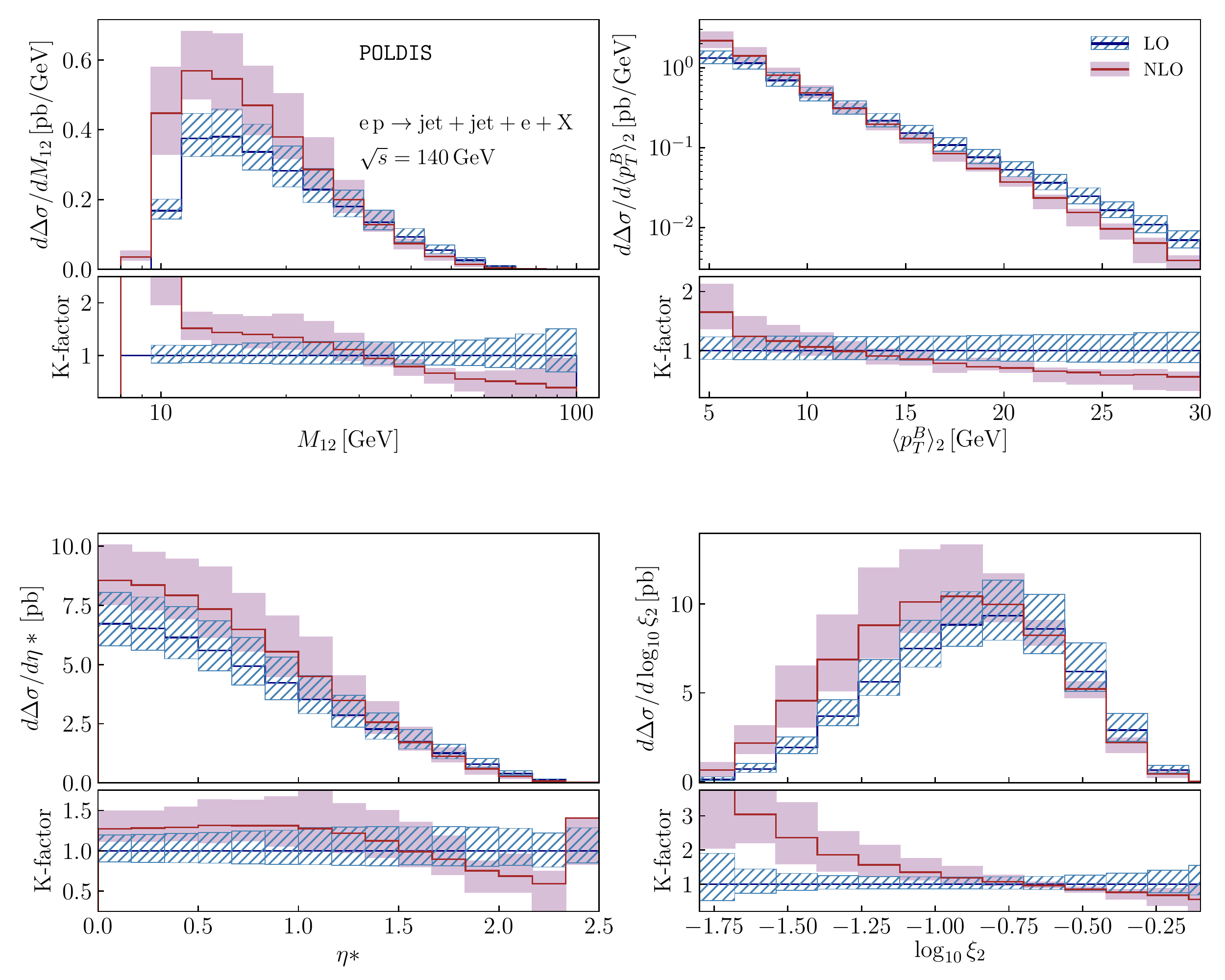, width=0.9\textwidth}
  \caption{Same as Fig. \ref{fig_dist_nopol}, but for the $polarized$ case.}\label{fig_dist_pol}
\end{figure}

While single-jet production can be described in terms of the jet pseudorapidity and transverse momentum, the availability of a second jet allows to define more kinematical observables to analyze the underlying partonic kinematics in detail. In that sense, it is instructive to study the unpolarized cross section as a function of the usual di-jet kinematical observables $\langle p_T^B \rangle_2$, $M_{12}$, $\eta^*$ and $\xi_2$, defined in section \ref{sec:kinematics}, as presented in Fig. \ref{fig_dist_nopol}. 

As it was noted for the kinematics of HERA \cite{Currie:2017tpe}, higher order corrections are sizable for all the variables under consideration. The scale variations of the NLO calculation are as large as the LO ones, or even larger, in the lower bins of the $M_{12}$, $\langle p_T^B \rangle_2$ and $\xi_2$ distributions, as the infrared limit is approached. As mentioned, this behaviour is mainly due to the asymmetrical cuts in $p_T$ imposed to the two jets. In the Breit frame, LO kinematics implies that the two outgoing partons generating the jets have opposite transverse momentum, and therefore the region with $\langle p_T^B \rangle_2<5$ GeV is not accessible at that order. A similar argument can be used to show that new regions of $M_{12}<10$ GeV and low $\xi_2$ become accessible only at NLO. This discrepancy in the available phase space at different orders is known to cause instabilities in the perturbative expansion \cite{Catani:1997xc}. Actually, for that {\it forbidden phase space region} the calculation is effectively a LO one. Note, however, that the use of symmetric cuts in $p_T$ leads to even worse perturbative problems, due to the enhancement of large logarithmic contributions related to the back-to-back configuration that can completely spoil the convergence of the expansion \cite{Frixione:1997ks,Currie:2017tpe}. The NLO corrections show a clear pattern, shifting the distributions to lower values of $\langle p_T^B \rangle_2<5$ GeV, which are in turn correlated to lower values of $M_{12}$ and $\xi_{2}$, and higher values of $\eta^{*}$.  

\begin{figure}
 \epsfig{figure= 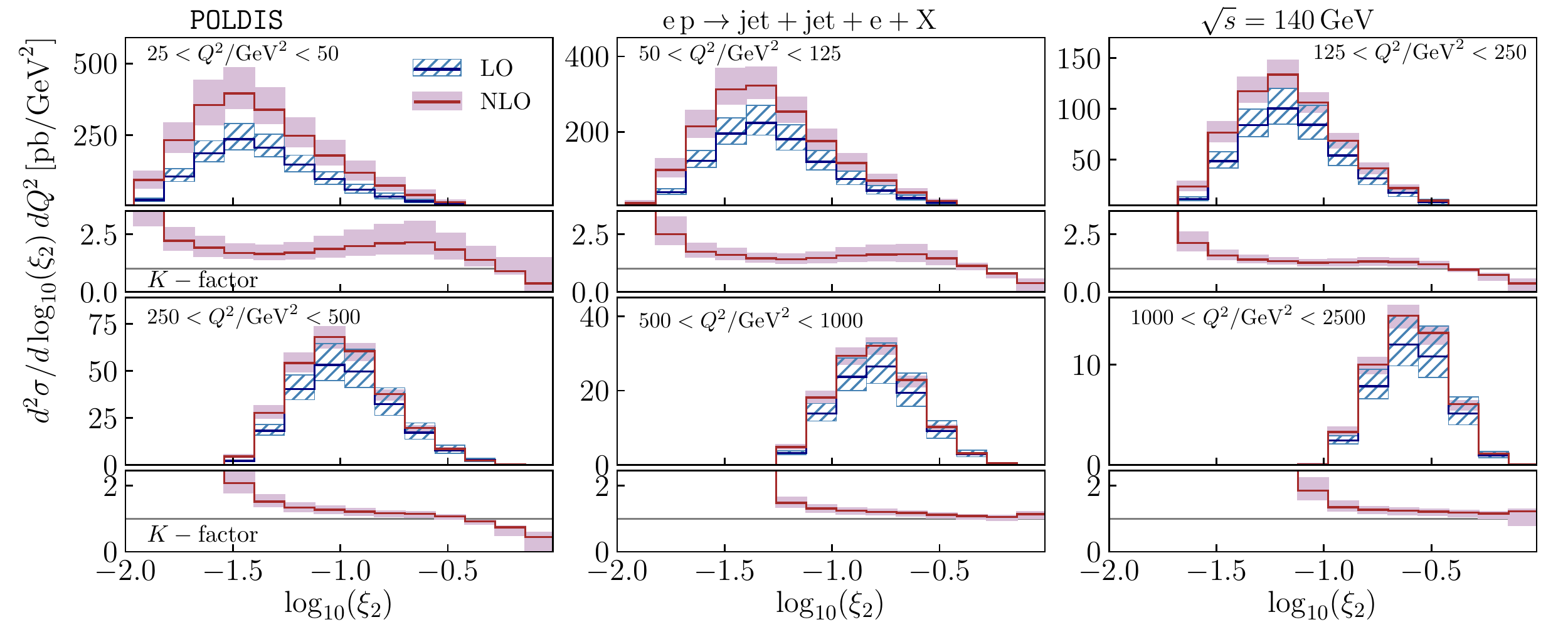 ,width=0.95\textwidth}
  \caption{Inclusive di-jet production unpolarized cross sections in bins of $Q^{2}$, as a function of $\log_{10}(\xi_2)$. The LO and NLO uncertainty bands are obtained as in Fig. \ref{fig_dist_pol}. The lower panels display the $K$-factors to the LO calculation}\label{fig_dobledist_log_nopol}
\end{figure}
\begin{figure}
 \epsfig{figure= 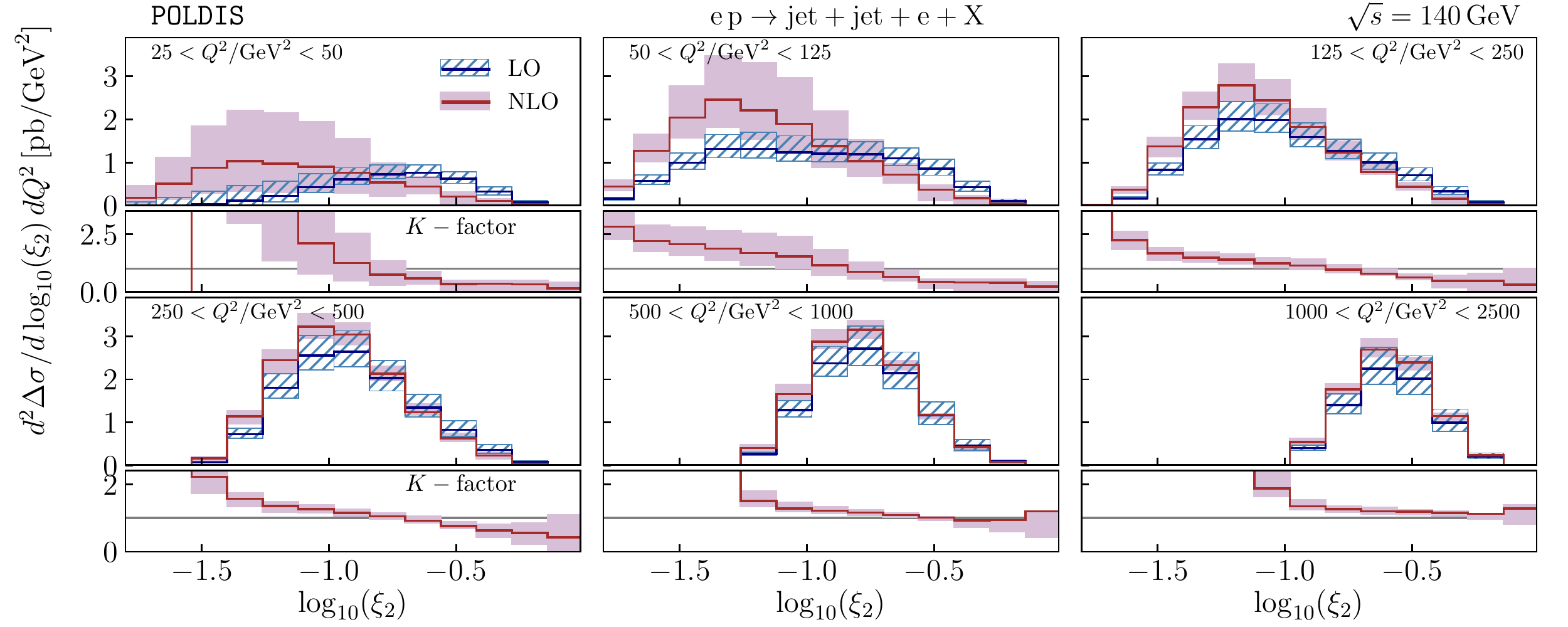 ,width=0.95\textwidth}
  \caption{Same as Fig. \ref{fig_dobledist_log_nopol}, but for the $polarized$ case.}\label{fig_dobledist_log_pol}
\end{figure}

In Fig. \ref{fig_dist_pol} we show the same distributions of Fig. \ref{fig_dist_nopol} but for the polarized cross section. Compared to the unpolarized case, for low $M_{12}$, $\langle p_T^B \rangle_2$, $\eta^*$ and $\xi_{2}$ it can be seen that while the NLO corrections follow the same pattern, they are generally milder, with lower $K$-factors. There is also a difference in the behaviour of the second order corrections for higher values of $M_{12}$, $\eta^*$ and $\xi_{2}$, resulting in stronger suppressions than the ones observed in the unpolarized case. The $\xi_{2}$ distribution is particularly shifted towards higher momentum fractions. The same considerations regarding theoretical uncertainties apply to the polarized case, leading to the strong NLO scale-dependence. 

\begin{figure}
 \epsfig{figure= 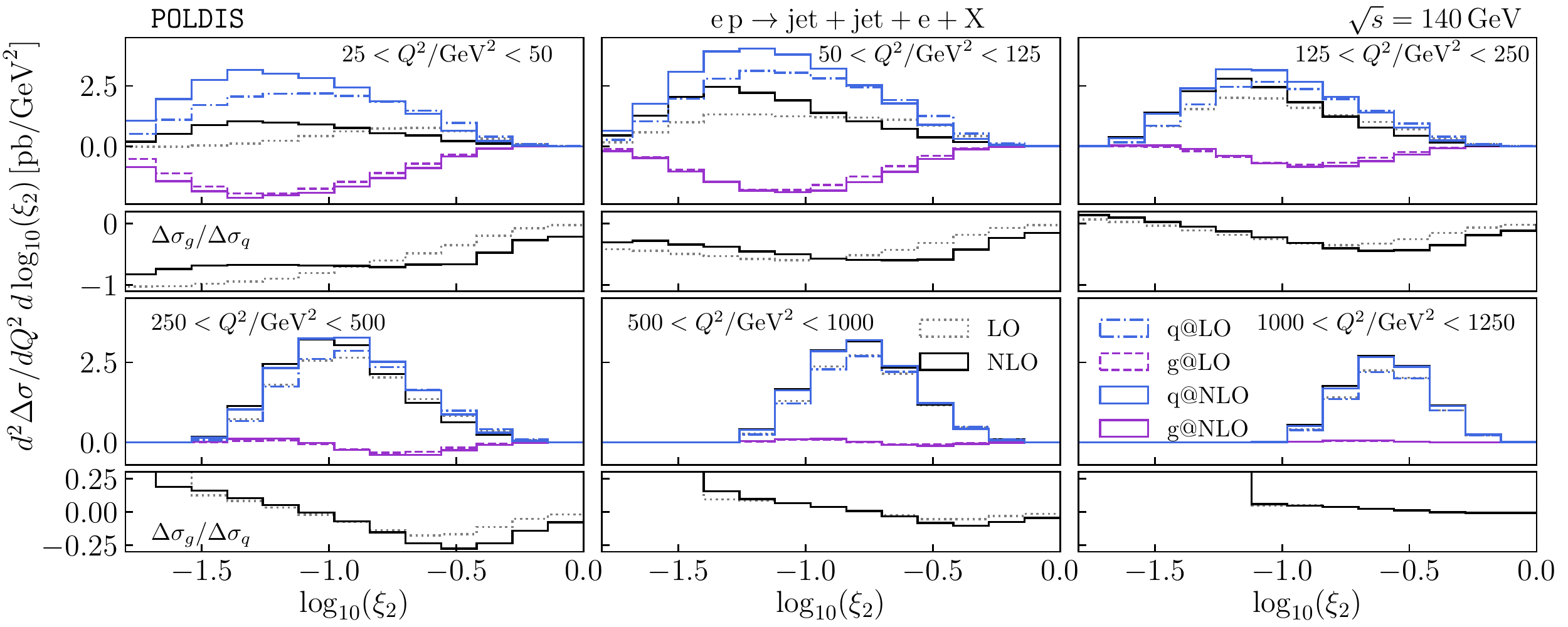 ,width=0.95\textwidth}
  \caption{Same as Fig. \ref{fig_dobledist_log_pol}, but separating the contributions initiated by quarks and gluons to the di-jet cross section, both at LO (dashed lines) and at NLO (solid lines). The lower insets show the ratio between the gluonic and quark contributions to the cross section.} \label{fig_dobledist_log_canales}
\end{figure}

The somewhat big NLO corrections, and the differences between the unpolarized and polarized cases, can be better understood by analysing the previous distribution at different values of $Q^2$. As an example, in Figs. \ref{fig_dobledist_log_nopol} and \ref{fig_dobledist_log_pol} we present the unpolarized and polarized double-differential distribution, i.e, in bins of $Q^{2}$ and $\log_{10} (\xi_{2})$, respectively. Regarding the unpolarized distributions of Fig.~\ref{fig_dobledist_log_nopol} it can be noted that, as expected, lower $Q^2$ values are correlated to smaller momentum fractions, from which the cross section receives its most important contributions. Di-jet production measurements at the EIC are therefore expected to explore the mid-$x$ region, $10^{-2}<x<10^{-1}$. The NLO cross sections for the high $Q^{2}$ bins are in good agreement with the LO calculations and show small scale dependence, indicating good convergence of the perturbative series. In addition to the complementary constraints on the quarks polarized and unpolarized distribution functions, restrictions coming from this region on the gluon helicity distribution, which is mainly probed down to $x \sim5\times10^{-2}$ by RHIC data, will be specially important. On the other hand, in Fig. \ref{fig_dobledist_log_nopol} it can be seen that both the $K$-factors and theoretical uncertainties increase as lower $Q^2$ values are considered. This is consistent with the aforementioned population of the new phase space region at low $\xi_{2}$ becoming available at NLO.

Compared to the unpolarized case, the polarized distributions of Fig. \ref{fig_dobledist_log_pol} present two striking features: they decrease at lower $Q^2$, and they also display significant differences in shape between LO and NLO results in that region. Both features can be explained by the analysis of the contributions from the quark and gluon channels to the polarized cross section. In Fig. \ref{fig_dobledist_log_canales} we present, precisely, the di-jet double-differential polarized distribution as a function of $Q^{2}$ and $\log_{10} (\xi_{2})$, distinguishing the contributions initiated by the quark and gluon channels. In this case, the lower insets in the plot depict the ratio between the gluon- and quark-initiated differential cross sections. The peculiar behaviour of the polarized cross sections as a function of $Q^{2}$ can be traced back to the negative sign of the gluon contribution below $Q^2=600$ GeV, which becomes more significant for lower values of $Q^{2}$, as shown in the ratio between the gluon and quark contributions. The enhancement of the negative contribution from the gluonic channel at low-$Q^{2}$ leads to strong cancellations against the positive quark contributions, and therefore to a reduction in the polarized cross section. At the same time, this reduction leads to larger relative theoretical uncertainties. The cancellation between channels is also responsible for the change in the behaviour observed in the $\xi_2$ distributions in Figs. \ref{fig_dist_pol} and \ref{fig_dobledist_log_pol} with respect to the unpolarized case, as well as the different shapes of the NLO corrections in the distributions of the other kinematical variables. Fig. \ref{fig_dobledist_log_canales} also shows that the NLO shift in the polarized $\xi_{2}$ distribution of Fig. \ref{fig_dist_pol} is mostly associated to a shift in the quark contribution towards lower momentum fractions.

\begin{figure}
 \epsfig{figure= 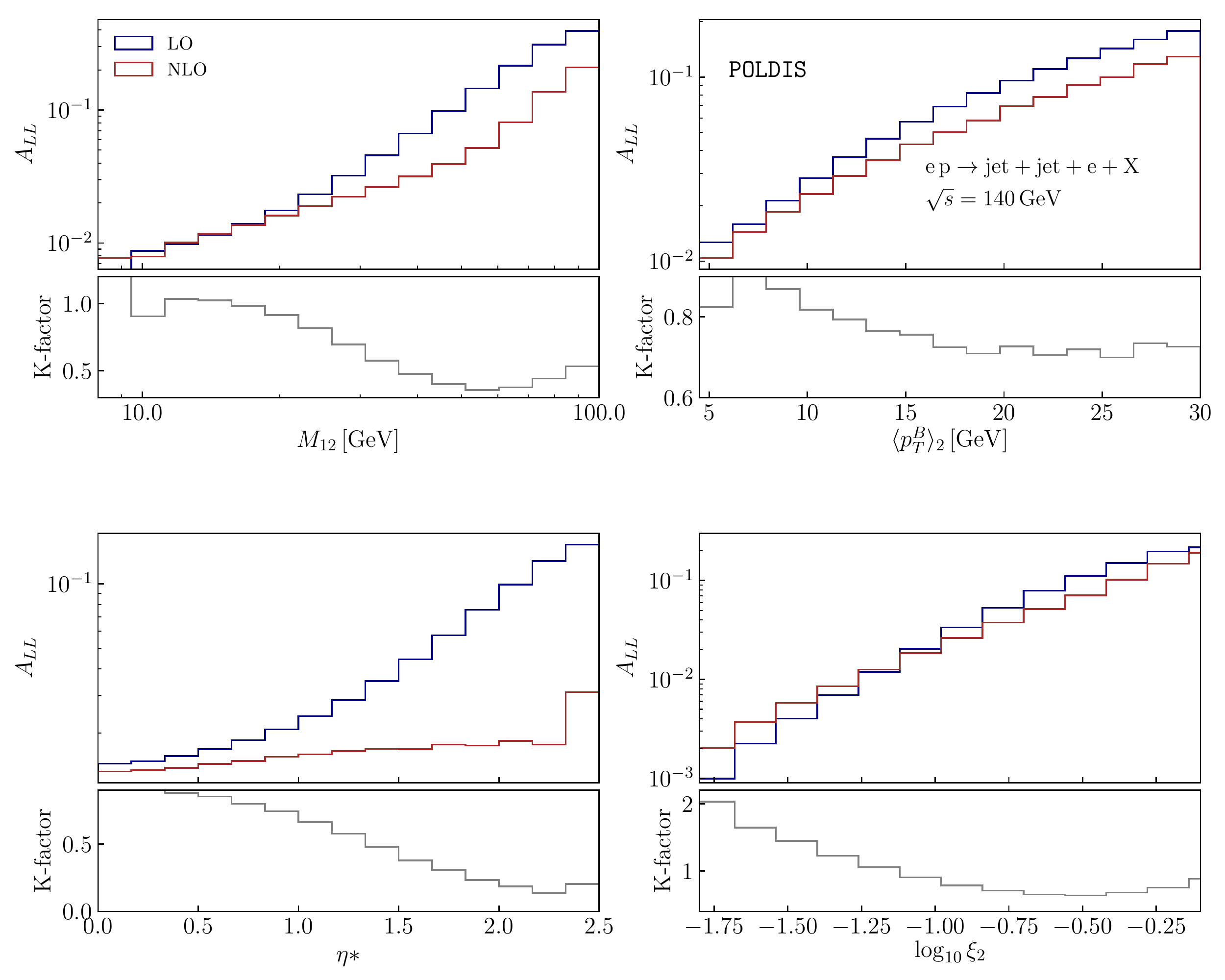 ,width=0.9\textwidth}
  \caption{Double spin asymmetries for di-jet production, as a function of $\langle p_T^B \rangle_2$, $M_{12}$, $\eta^*$ and $\xi_2$. The lower boxes depict the $K$-factors to the LO calculation.}\label{fig_asymmetry}
\end{figure}

\begin{figure}
 \epsfig{figure= 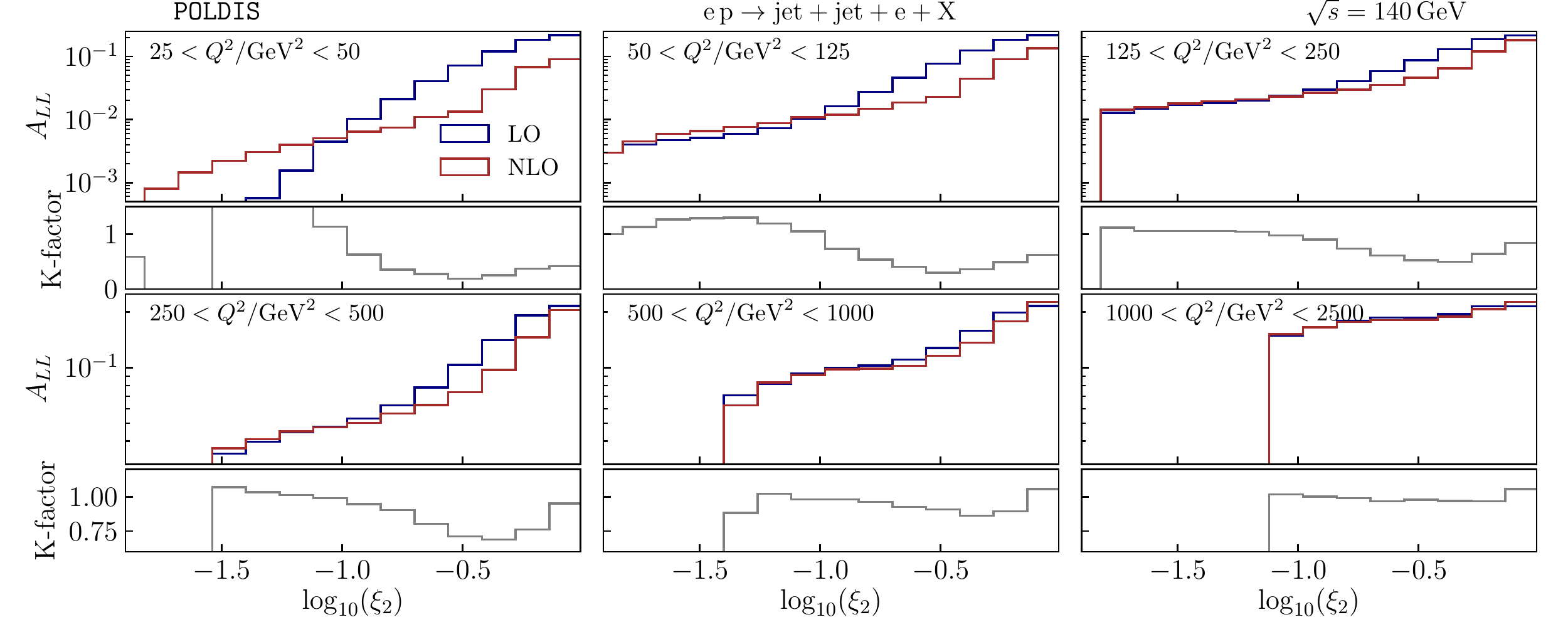 ,width=0.95\textwidth}
  \caption{Double spin asymmetry as a function of $\log_{10}(\xi_{2})$ and $Q^{2}$. As in the previous cases, the inset lower boxes show the $K$-factor to the LO asymmetry.}\label{fig_dobledist_asymmetry}
\end{figure}

The most relevant observables in polarized processes are the double spin asymmetries, defined as the ratio of the polarized and unpolarized cross sections $A_{LL}=\Delta \sigma / \sigma$, since the cancellation of systematic uncertainties is expected to happen in the quotient. In Fig. \ref{fig_asymmetry} we show the double spin asymmetries for $\langle p_T^B \rangle_2$, $M_{12}$, $\eta^*$ and $\xi_2$. The asymmetries values are typically of order $\sim(1-10 \%)$ in the relevant regions, with a significant reduction in the NLO values due to the higher $K$-factors observed in the unpolarized cross sections. The only exception is the $\xi_2$ distribution, where the unusual behaviour due to gluon cancellation and the shift in the quark contribution to the polarized cross section leads to an enhancement in the asymmetry at lower momentum fractions, albeit the very small values of the asymmetry in that region. 

Once again, the behaviour of the asymmetries can be better understood by studying the double-differential $Q^2$ dependence of the distributions. Fig. \ref{fig_dobledist_asymmetry} depicts the double spin asymmetry as a function of both $Q^{2}$ and $\xi_2$. The reduction of the polarized cross sections for low values of $Q^{2}$ due to the negative gluonic contribution leads to a sizable suppression of the asymmetry in those bins for $\xi_2\gtrsim 10^{-1}$. It is worth mentioning that, for the first to bins of $Q^{2}$, the significant shift in the NLO quark contribution towards lower momentum fractions shown in Fig.~\ref{fig_dobledist_log_canales} results in an enhancement of the asymmetries for $\xi_2\lesssim 10^{-1}$ .The clear pattern of the NLO corrections to the asymmetries can be easily understood by the direct comparison of the Figs. \ref{fig_dobledist_log_nopol} and \ref{fig_dobledist_log_pol}: high $Q^{2}$ bins show $K$-factors close to 1, due to the smaller corrections observed for both the polarized and unpolarized distributions. The difference in sign of the NLO corrections for each case in the high momentum fraction region results in the reduction of the asymmetries shown in Fig. \ref{fig_dobledist_asymmetry}, which becomes more important as lower values of $Q^{2}$ are reached. This very same behaviour has been seen for the other kinematical observables $M_{12}$, $\langle p_T^B \rangle_2$ and $\eta^*$.

\section{Polarized NNLO inclusive-jet Production}\label{sec:single-jets}

Having discussed our NLO di-jet production calculation, we can now turn to the NNLO corrections for single jet production, obtained through the application of the P2B method. In this section, we present our results for polarized single-inclusive jet production at NNLO in the laboratory frame (L), for the Electron-Ion-Collider kinematics. Similarly to \cite{Borsa:2020ulb}, the default distributions are obtained reconstructing the jets with the anti-$k_{T}$ algorithm and $E_{T}$-scheme recombination, using a jet radius $R=0.8$, and fixing the normalization and factorization scales central values as $\mu_{F}^{2}=\mu_{R}^{2}=Q^{2}\equiv\mu_0$. As in the previous section, $\alpha_s$ is evaluated at NLO accuracy with $\alpha_s(M_z)=0.118$. The reconstructed jet in the laboratory frame is then required to satisfy:

\begin{equation}
\begin{tabular}{ c }
    $5\,\mathrm{GeV}<p^{L}_{T}<36\,\mathrm{GeV}$,\\
    $|\eta^{L}|<3$,
\end{tabular}
\end{equation}

\noindent while on the leptonic side we impose the additional cuts:

\begin{equation}
\begin{tabular}{ c }
    $0.04<y<0.95$,\\
    $25\, \mathrm{GeV}^{2} <Q^{2}<1000\, \mathrm{GeV}^{2}$.
\end{tabular}
\end{equation}

 \noindent The lower cut in $Q^{2}$ was chosen to avoid differences in the phase space available at different orders. Note that at LO the transverse momentum of the jet in the laboratory frame is given by $(p^{L}_{T})^2=Q^{2}\,(1-y)$, and thus the region $Q^{2}\lesssim 25$ GeV$^{2}$ is kinematically forbidden for the specified cuts in $p^L_{T}$. Since there is no NNLO global fit of polarized PDFs available, the parton distributions sets used were, once again, the NLO extractions NLOPDF4LHC15 \cite{Butterworth:2015oua} and DSSV \cite{deFlorian:2014yva,deFlorian:2019zkl} for the unpolarized and polarized case, respectively.

\begin{figure}
 \epsfig{figure= 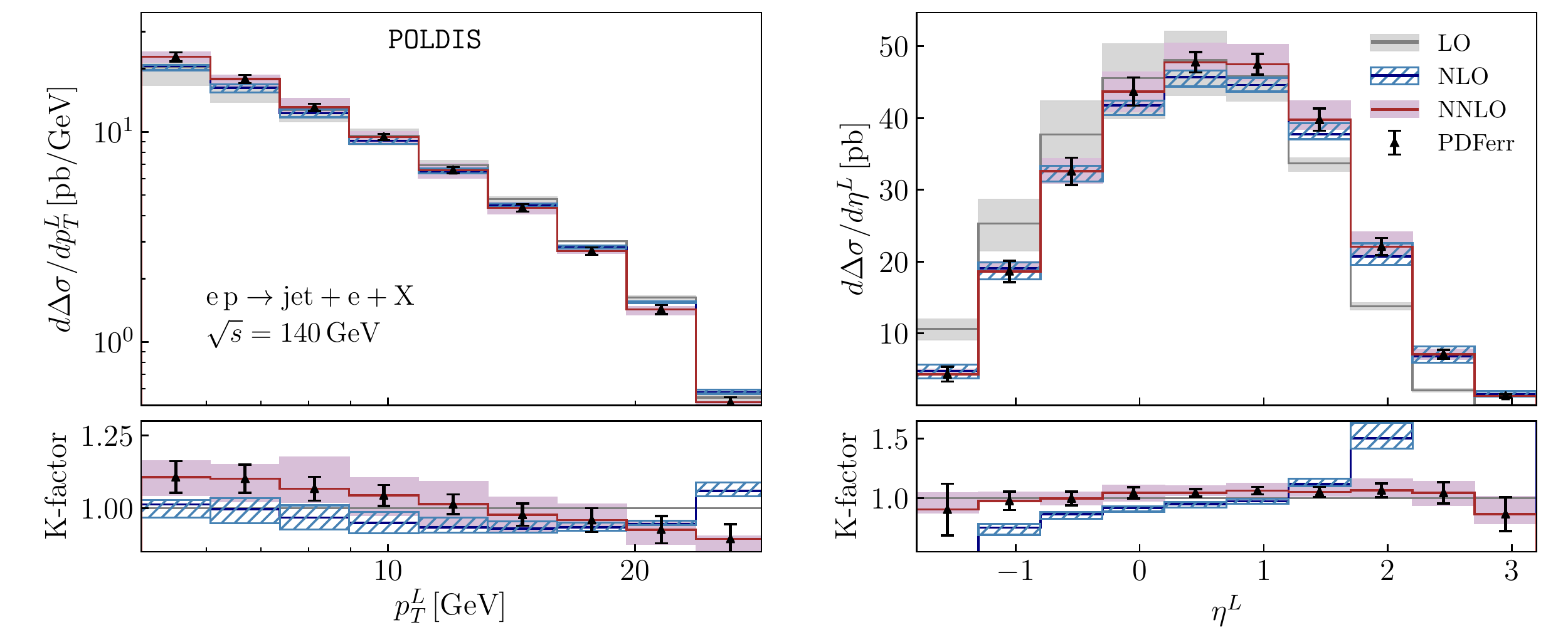, width=0.95\textwidth}
 \epsfig{figure= 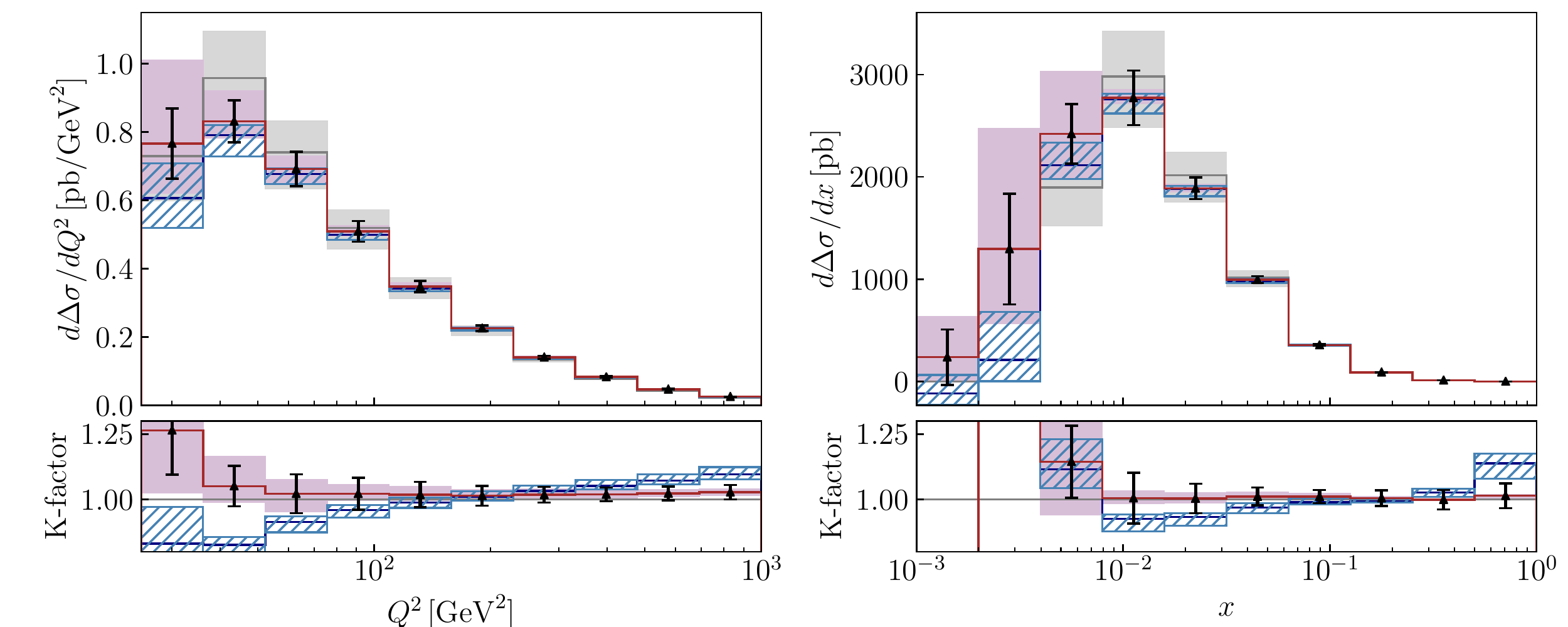, width=0.95\textwidth}
  \caption{Polarized single-jet cross section, as distributions in transverse momentum $p^L_{T}$, pseudorapity $\eta$, photon virtuality $Q^2$ and Bjorken variable $x$ at LO, NLO and NNLO. The bands reflect the seven point variation in the cross section when independently changing the scales as $\mu_R,\mu_F= [1/2,2]\mu_0$. The error bars correspond to the PDF's errors in the NNLO result. The lower inset shows the corresponding K-factors, as defined in the main text.}\label{fig:singlejet_dist_pol}
\end{figure}

In Fig. \ref{fig:singlejet_dist_pol} we present the cross section for single-inclusive jet production in polarized DIS, as a function of the jet transverse momentum $p^L_{T}$, its pseudorapidity $\eta^L$, and in terms of $Q^2$ and $x$, calculated at LO, NLO and NNLO accuracy. The lower insets in Fig. \ref{fig:singlejet_dist_pol} show the K-factors, defined as the ratios to the previous order, that is, $K^{\mathrm{NNLO}}=\sigma^{\mathrm{NNLO}}/\sigma^{\mathrm{NLO}}$ and $K^{\mathrm{NLO}}=\sigma^{\mathrm{NLO}}/\sigma^{\mathrm{LO}}$. As in the case of di-jet production, the theoretical uncertainty bands were obtained performing a seven-point independent variation of the renormalization and factorization scales as $\mu_{R}, \mu_{F}=[\frac{1}{2},2]\mu_0$. The uncertainty associated to the polarized parton distributions was estimated using the DSSV set of PDFs replicas from \cite{deFlorian:2019zkl}. Note that due to the unavailability of proper polarized NNLO PDF, this bands should be taken only as a first attempt to quantify the non-perturbative errors in the NNLO cross section. The same NLO PDFs were used at all orders so as to quantify only the variations arising from the perturbative calculation.

As it can be seen in Fig. \ref{fig:singlejet_dist_pol}, the main effect of higher order corrections is to \textit{shift} these distributions toward higher values of pseudorapidity and lower values of transverse momentum, since more jets originating from the emission of additional partons become available in those regions. In the case of the pseudorapidity distribution, this is translated into high values of the NLO K-factor in the forward region ($\eta^L>1$), while a strong suppression in the backward region ($\eta^L<-1$) is observed. NNLO corrections have the same behaviour, albeit with lower values of K-factor. Similar comments can be made regarding the transverse momentum distribution, which is enhanced for lower values of $p^L_{T}$. 

For the $p_{T}$ distribution, the NNLO corrections are typically of order $10\%$, while for the $\eta$ distributions they are of order $5\%$. It should be noted that while there is good agreement between the NLO and NNLO calculations, with overlapping bands throughout the kinematical range, anticipating convergence of the perturbative series, the scale bands for the NNLO distributions are still somewhat large in certain bins compared with those of the NLO. This effect is associated with the kinematical suppression of the LO contributions in some regions due to the cut enforced in $y$ and $p^L_{T}$, which spoils the accuracy of the perturbative series in that region. This can be better observed in the $Q^2$ and $x$ distributions in Fig. \ref{fig:singlejet_dist_pol}. At low values of $Q^2$ and $x$ the suppression of the Born cross section clearly correlates with higher K-factors and scale bands, especially for $x\lesssim 5\times10^{-3}$, where the LO is completely forbidden. The same effect is present in the unpolarized case for the aforementioned regions. 

\begin{figure}
\epsfig{figure= 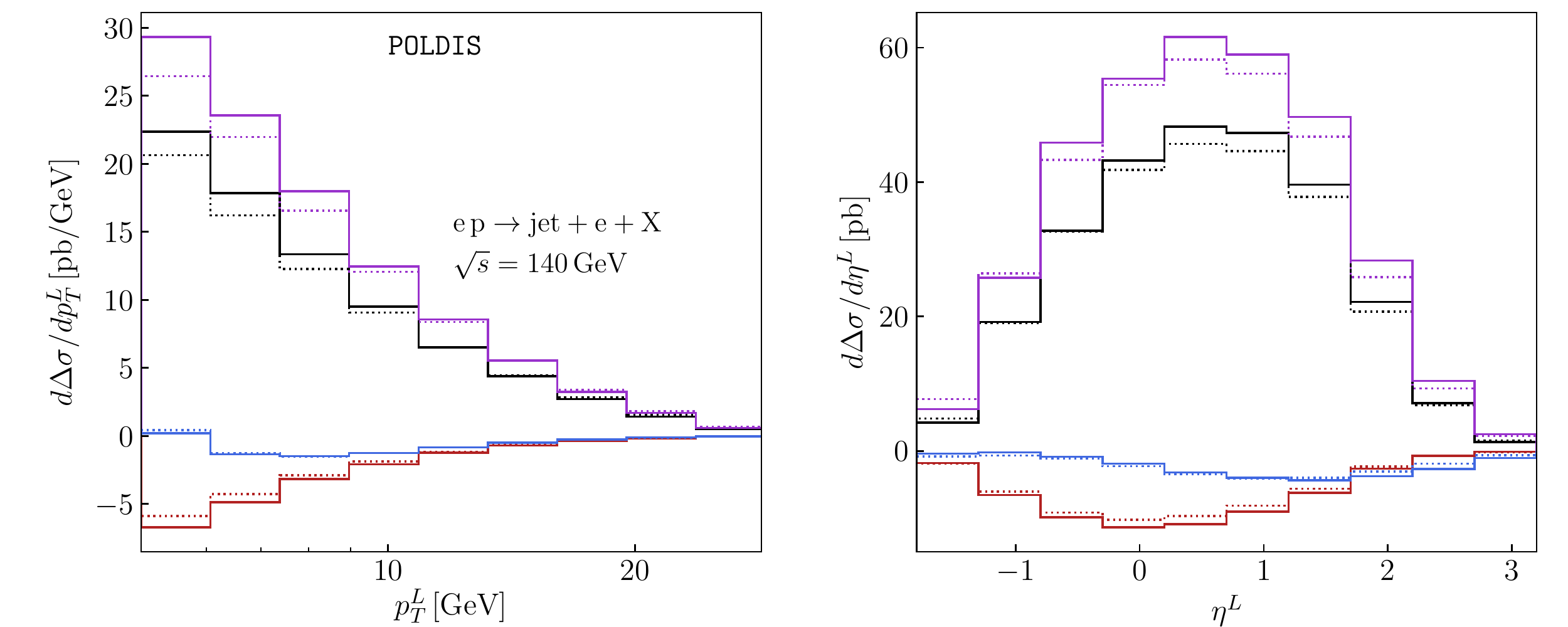, width=0.95\textwidth}
\epsfig{figure= 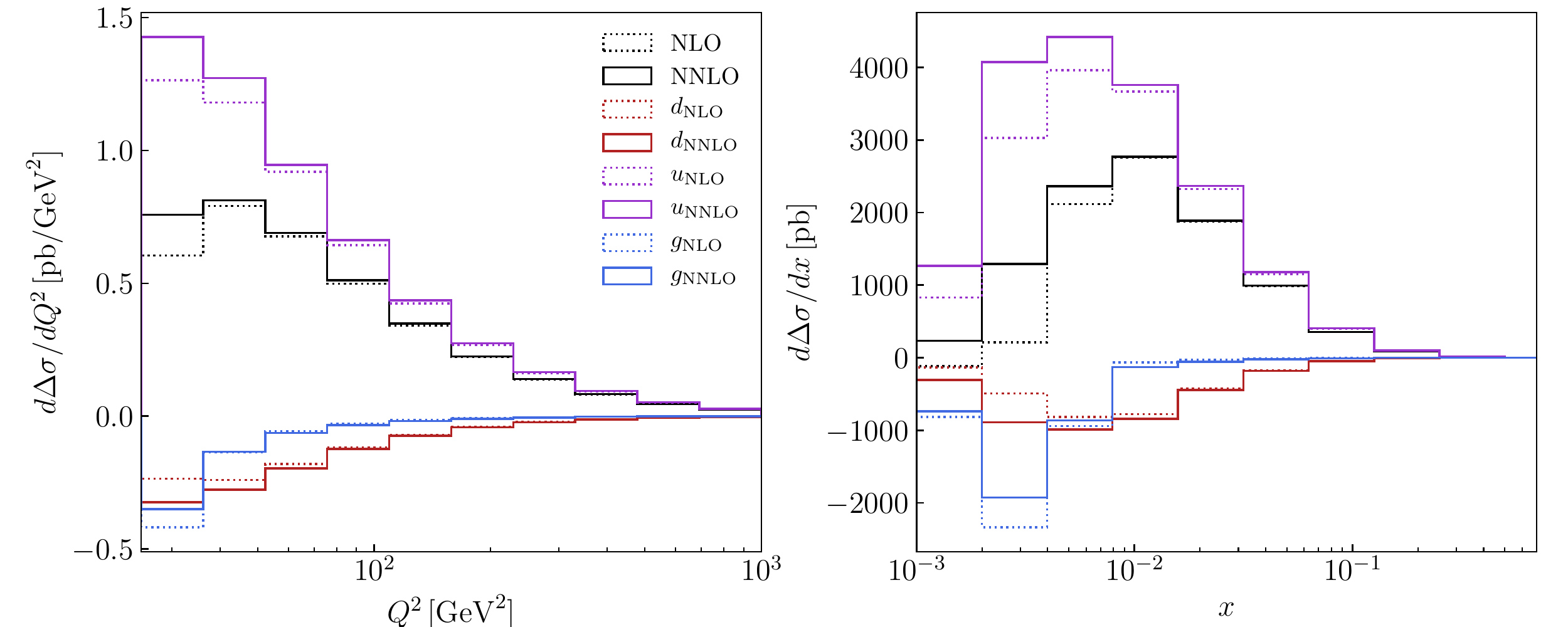, width=0.95\textwidth}
 \caption{Same as Fig. \ref{fig:singlejet_dist_pol}, but discriminating the contributions to the cross section coming from initial quarks and gluons.}\label{fig:singlejet_canales_pol}
\end{figure}

Even though the growth of the uncertainty bands at NNLO in the $p^L_T$ and $\eta$ distributions originates from the difference in the available phase space at each order, the sizes of the bands in this region are further enhanced in the polarized case compared to the unpolarized one. This results in bigger NNLO bands in $p^L_T$ and $\eta^L$ distributions, as observed in \cite{Borsa:2020ulb}. This enhancement is related to the fact that in the polarized case there are cancellations between processes initiated by different partons. To highlight this point, in Fig. \ref{fig:singlejet_canales_pol} we present the contributions of the most relevant parton channels to the polarized cross section. As in the unpolarized case, for most of the explored $Q^{2}$ and $x$ values, the cross section is dominated by initial $u$ quark contributions. However, as lower values of both $Q^2$ and $x$ are reached, there are significant cancellations between the $u$ quark channel and the negative contribution of the $d$ quark and gluon channels, which accounts for higher relative uncertainties once the sum over of all the initial parton contributions is taken (the $s$ quark also has a negative contribution, but it is negligible). Since low $Q^{2}$ and $x$ correlate with low $p^L_{T}$ and $\eta^L \gtrsim 0$, those same cancellations are translated into the sizable NNLO scale bands in Fig. \ref{fig:singlejet_dist_pol} in those ranges. 

It is worth noticing that, even though it is expected to have a greater gluon contribution at low $p^L_T$, since that region correlates with low $Q^{2}$, the first bin of the $p^L_T$ distribution is very small and slightly positive (as opposed to the $u$ and $d$ quarks contributions). This is related to the fact that the gluon contribution to the structure function is positive below $x \sim 2 \times 10^{-2}$. Since the structure function is obtained by the integration over all the $p^L_T$ range, as lower values of $p^L_T$ are reached the $p^L_T$ distribution must become positive at some point. 

Regarding the uncertainty associated to the PDFs, it is typically of order $5\%-10\%$ for the region of $\{p^L_{T},\eta^L\}$ studied. Though this uncertainty is comparable to the NNLO corrections for most of the kinematical range, it should be noted that for the low $p^L_{T}$ region, it becomes smaller than the NNLO corrections, highlighting the relevance that NNLO extractions will have in order to match the accuracy of the perturbative side. As in the case of the scale-variations bands, the PDF uncertainty becomes larger as lower values of $x$ and $Q^{2}$ are approached, since the cancellation between the different partonic channels for those bins is sensitive to changes in the partonic distributions. 

\begin{figure}
 \epsfig{figure= 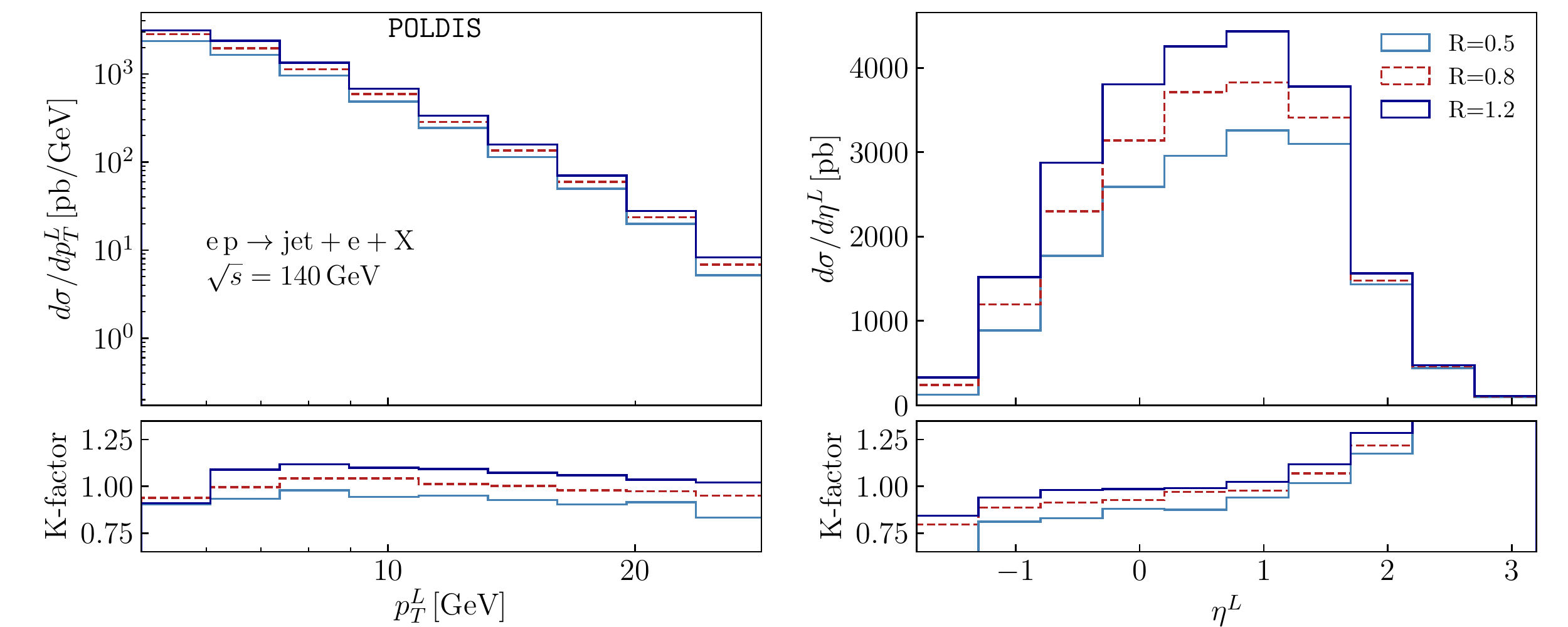, width=0.95\textwidth}
 \epsfig{figure= 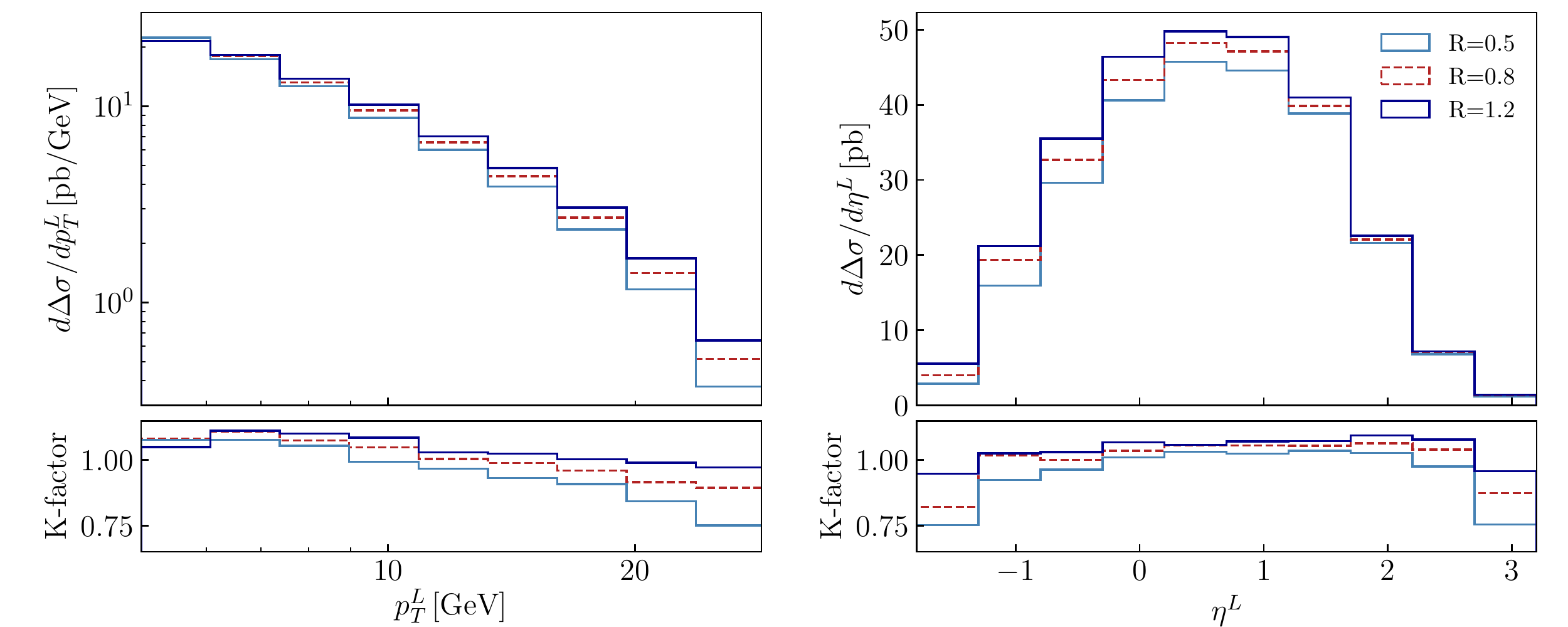, width=0.95\textwidth}
  \caption{NNLO polarized and unpolarized cross sections as a distribution of $p^L_{T}$ and $\eta^L$ for different choices of the jet radius $R$ value.}\label{fig:singlejet_R_pol}
\end{figure}

Another feature associated to cancellation between partonic channels in the polarized cross section is the reduced dependence on the parameters of the jet-reconstruction algorithm, compared to the unpolarized case. To emphasize this point, in Fig. \ref{fig:singlejet_R_pol}, we present the NNLO cross sections as a distribution of both $p^L_{T}$ and $\eta^L$, for different values of the jet radius $R=0.5, 0.8, 1$ used in the anti-$k_{T}$ algorithm. In both cases, higher values of jet radius correspond to larger cross sections in the whole kinematical range due to the inclusion of more jets that satisfy the imposed cuts. However, the polarized case shows a reduced dependence in $R$ at low $p^L_T$ and the intermediate $\eta^L$ values, precisely where the stronger cancellations between channels take place. This results in an overall reduction of the dependence of the polarized cross section on the jet parameter. It is worth noticing that while the total cross section is affected by these strong cancellations between channels, with the use of jet tagging techniques~\cite{Arratia:2020azl,Kang:2020fka} it could be possible to noticeably modify the shape of the distributions, enhancing the contributions from different partons.

\begin{figure}
 \epsfig{figure= 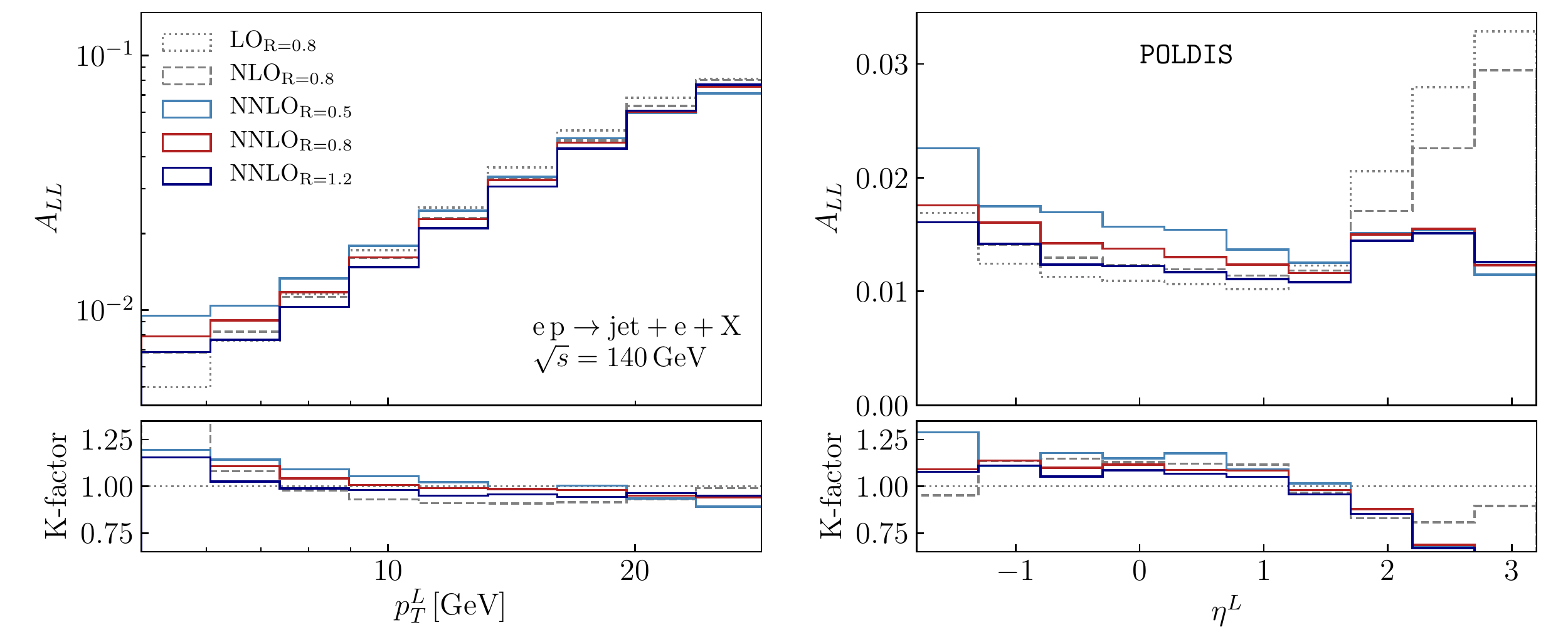, width=0.95\textwidth}
  \caption{NNLO double spin asymmetries as a distribution of $p^L_{T}$ and $\eta$ for different choices of the jet radius $R$ value.}\label{fig:asymm_R}
\end{figure}

The difference of sensitivity to changes in the jet radius will in turn modify the behaviour of the double spin asymmetries. In Fig. \ref{fig:asymm_R} we present the NNLO double spin asymmetries in the $p^L_T$ and $\eta^L$ distributions for the $R$ values used before. As expected, a larger dependence on $R$ is obtained in those regions where the cancellation between channels for the polarized cross sections are more important. For those regions, the increase in $R$ leads to a relative increase of the unpolarized cross section, and consequently to a reduction in the spin asymmetry. Conversely, lower values of $R$ produce an increment of the asymmetry in the same regions. Fig. \ref{fig:asymm_R} also shows the LO and NLO asymmetries for $R=0.8$. Regarding higher order corrections, it is worth mentioning that the relative higher NNLO contributions to unpolarized cross section lead to an important suppression of the asymmetry in the high pseudorapidity region, with milder corrections for intermediate $\eta^L$. However, note that for $\eta^L \lesssim 1$ and $p_{T}\lesssim 10$, the variations with the jet radius are greater than those coming from the perturbative series. The jet parameters are therefore expected to have sizable impact in the double spin asymmetries in regions where cancellation between partonic contributions take place in the polarized cross section.

\section{Conclusions}\label{sec:conclusion}

In this paper we have presented the NLO calculation for the production of di-jets in polarized and unpolarized lepton-nucleon DIS in the Breit frame, for the EIC kinematics. Our calculation is based in a generalization of the dipole subtraction method to handle the polarization of initial-state particles, which is discussed in detail. The cross sections were studied as functions of the leading jets transverse momenta $p^B_{T,1}$ and $p^B_{T,2}$, invariant mass of the jets $M_{12}$, the mean transverse momentum $\langle p^B_{T}\rangle_{2}$, the difference in pseudorapidities $\eta^{*}$ and the di-jet momentum fraction $\xi_2$. Additionally, the double-differential distributions in $Q^{2}$ and $\xi_{2}$ were analyzed. Both for the polarized and unpolarized cross sections, the differential distributions show important NLO corrections, particularly for low values of $M_{12}$ and $\xi_2$, and higher values of $\eta^{*}$, associated to differences in the phase space available at each order. While the NLO corrections obtained show good agreement with the LO calculations and reduced dependence on the choice for the factorization and renormalization scales, for values of $Q^{2}$ above 250 GeV, anticipating convergence of the perturbative expansion, the distributions for lower values of $Q^{2}$ present sizable corrections as well as a strong dependence on the scale choice. We noted that this effect is further enhanced in the polarized cross sections, due to the non-negligible negative contribution of the gluon-initiated channel, producing noticeable differences between the polarized results and their unpolarized counterparts. This difference in behaviour is translated to the double spin asymmetries, with significant suppression in $M_{12}$, $\eta^{*}$ and $\langle p^B_{T}\rangle_{2}$. Once again, the corrections are more significant as lower values values of $Q^{2}$ are approached.

The di-jet calculation was in turn used to obtain the polarized NNLO single-inclusive jet production cross-section in the laboratory frame via the P2B method~\cite{Borsa:2020ulb}, which combines the exclusive NLO di-jet cross section along with the inclusive NNLO polarized structure function. We expanded on our previous results to include a better estimate of the theoretical uncertainty, as well as the dependence on the jet radius. Good agreement was found between the NLO and NNLO results for the range studied in $p^L_{T}$ and $\eta^L$. The somewhat large size of some of the NNLO uncertainty bands was linked to a combination of the effects due to the difference in phase space available at LO at low $Q^2$ and $x$, also present in the unpolarized case, as well as the cancellation between partonic channels in the polarized cross section. This channel cancellation also leads to a reduced dependence of the polarized cross section in the jet radius $R$, which in turn produces a more noticeable dependence of the double spin asymmetries in $R$ in the regions of low $p_T^L$ and intermediate values of $\eta^L$. This hints towards a sizable dependence of the polarized cross section and asymmetries with the jet parameters in those regions, as well as important sensibility to the recently proposed jet-tagging techniques.

The results presented on this paper highlight the relevance that higher order QCD corrections will have in the precise description of the jet observables to be obtained in the future EIC, as well as the potential of those measurements to further improve our understanding of the spin structure of the proton and, particularly, in the precise extraction of polarized parton distributions.

\acknowledgments
We thank Rodolfo Sassot for discussions and Stefano Catani, Gavin Salam and Mike Seymour for useful communications.
This work was partially supported by CONICET and ANPCyT.

\bibliography{refs}

\begin{thebibliography}{48}
\expandafter\ifx\csname natexlab\endcsname\relax\def\natexlab#1{#1}\fi
\expandafter\ifx\csname bibnamefont\endcsname\relax
  \def\bibnamefont#1{#1}\fi
\expandafter\ifx\csname bibfnamefont\endcsname\relax
  \def\bibfnamefont#1{#1}\fi
\expandafter\ifx\csname citenamefont\endcsname\relax
  \def\citenamefont#1{#1}\fi
\expandafter\ifx\csname url\endcsname\relax
  \def\url#1{\texttt{#1}}\fi
\expandafter\ifx\csname urlprefix\endcsname\relax\def\urlprefix{URL }\fi
\providecommand{\bibinfo}[2]{#2}
\providecommand{\eprint}[2][]{\url{#2}}

\bibitem[{\citenamefont{Aidala et~al.}(2013)\citenamefont{Aidala, Bass, Hasch,
  and Mallot}}]{Aidala:2012mv}
\bibinfo{author}{\bibfnamefont{C.~A.} \bibnamefont{Aidala}},
  \bibinfo{author}{\bibfnamefont{S.~D.} \bibnamefont{Bass}},
  \bibinfo{author}{\bibfnamefont{D.}~\bibnamefont{Hasch}}, \bibnamefont{and}
  \bibinfo{author}{\bibfnamefont{G.~K.} \bibnamefont{Mallot}},
  \bibinfo{journal}{Rev. Mod. Phys.} \textbf{\bibinfo{volume}{85}},
  \bibinfo{pages}{655} (\bibinfo{year}{2013}), \eprint{1209.2803}.

\bibitem[{\citenamefont{Aschenauer et~al.}(2013)\citenamefont{Aschenauer,
  Bazilevsky, Boyle, Eyser, Fatemi et~al.}}]{Aschenauer:2013woa}
\bibinfo{author}{\bibfnamefont{E.}~\bibnamefont{Aschenauer}},
  \bibinfo{author}{\bibfnamefont{A.}~\bibnamefont{Bazilevsky}},
  \bibinfo{author}{\bibfnamefont{K.}~\bibnamefont{Boyle}},
  \bibinfo{author}{\bibfnamefont{K.}~\bibnamefont{Eyser}},
  \bibinfo{author}{\bibfnamefont{R.}~\bibnamefont{Fatemi}},
  \bibnamefont{et~al.} (\bibinfo{year}{2013}), \bibinfo{note}{arXiv:1304.0079}.

\bibitem[{\citenamefont{de~Florian et~al.}(2014)\citenamefont{de~Florian,
  Sassot, Stratmann, and Vogelsang}}]{deFlorian:2014yva}
\bibinfo{author}{\bibfnamefont{D.}~\bibnamefont{de~Florian}},
  \bibinfo{author}{\bibfnamefont{R.}~\bibnamefont{Sassot}},
  \bibinfo{author}{\bibfnamefont{M.}~\bibnamefont{Stratmann}},
  \bibnamefont{and}
  \bibinfo{author}{\bibfnamefont{W.}~\bibnamefont{Vogelsang}},
  \bibinfo{journal}{Phys. Rev. Lett.} \textbf{\bibinfo{volume}{113}},
  \bibinfo{pages}{012001} (\bibinfo{year}{2014}), \eprint{1404.4293}.

\bibitem[{\citenamefont{de~Florian et~al.}(2009)\citenamefont{de~Florian,
  Sassot, Stratmann, and Vogelsang}}]{deFlorian:2009vb}
\bibinfo{author}{\bibfnamefont{D.}~\bibnamefont{de~Florian}},
  \bibinfo{author}{\bibfnamefont{R.}~\bibnamefont{Sassot}},
  \bibinfo{author}{\bibfnamefont{M.}~\bibnamefont{Stratmann}},
  \bibnamefont{and}
  \bibinfo{author}{\bibfnamefont{W.}~\bibnamefont{Vogelsang}},
  \bibinfo{journal}{Phys. Rev. D} \textbf{\bibinfo{volume}{80}},
  \bibinfo{pages}{034030} (\bibinfo{year}{2009}), \eprint{0904.3821}.

\bibitem[{\citenamefont{Nocera et~al.}(2014)\citenamefont{Nocera, Ball, Forte,
  Ridolfi, and Rojo}}]{Nocera:2014gqa}
\bibinfo{author}{\bibfnamefont{E.~R.} \bibnamefont{Nocera}},
  \bibinfo{author}{\bibfnamefont{R.~D.} \bibnamefont{Ball}},
  \bibinfo{author}{\bibfnamefont{S.}~\bibnamefont{Forte}},
  \bibinfo{author}{\bibfnamefont{G.}~\bibnamefont{Ridolfi}}, \bibnamefont{and}
  \bibinfo{author}{\bibfnamefont{J.}~\bibnamefont{Rojo}}
  (\bibinfo{collaboration}{NNPDF}), \bibinfo{journal}{Nucl. Phys. B}
  \textbf{\bibinfo{volume}{887}}, \bibinfo{pages}{276} (\bibinfo{year}{2014}),
  \eprint{1406.5539}.

\bibitem[{\citenamefont{Accardi et~al.}(2016)}]{Accardi:2012qut}
\bibinfo{author}{\bibfnamefont{A.}~\bibnamefont{Accardi}} \bibnamefont{et~al.},
  \bibinfo{journal}{Eur. Phys. J. A} \textbf{\bibinfo{volume}{52}},
  \bibinfo{pages}{268} (\bibinfo{year}{2016}), \eprint{1212.1701}.

\bibitem[{\citenamefont{Aschenauer et~al.}(2012)\citenamefont{Aschenauer,
  Sassot, and Stratmann}}]{Aschenauer:2012ve}
\bibinfo{author}{\bibfnamefont{E.~C.} \bibnamefont{Aschenauer}},
  \bibinfo{author}{\bibfnamefont{R.}~\bibnamefont{Sassot}}, \bibnamefont{and}
  \bibinfo{author}{\bibfnamefont{M.}~\bibnamefont{Stratmann}},
  \bibinfo{journal}{Phys. Rev. D} \textbf{\bibinfo{volume}{86}},
  \bibinfo{pages}{054020} (\bibinfo{year}{2012}), \eprint{1206.6014}.

\bibitem[{\citenamefont{Aschenauer et~al.}(2015)\citenamefont{Aschenauer,
  Sassot, and Stratmann}}]{Aschenauer:2015ata}
\bibinfo{author}{\bibfnamefont{E.~C.} \bibnamefont{Aschenauer}},
  \bibinfo{author}{\bibfnamefont{R.}~\bibnamefont{Sassot}}, \bibnamefont{and}
  \bibinfo{author}{\bibfnamefont{M.}~\bibnamefont{Stratmann}},
  \bibinfo{journal}{Phys. Rev. D} \textbf{\bibinfo{volume}{92}},
  \bibinfo{pages}{094030} (\bibinfo{year}{2015}), \eprint{1509.06489}.

\bibitem[{\citenamefont{Aschenauer et~al.}(2020)\citenamefont{Aschenauer,
  Borsa, Lucero, Nunes, and Sassot}}]{Aschenauer:2020pdk}
\bibinfo{author}{\bibfnamefont{E.~C.} \bibnamefont{Aschenauer}},
  \bibinfo{author}{\bibfnamefont{I.}~\bibnamefont{Borsa}},
  \bibinfo{author}{\bibfnamefont{G.}~\bibnamefont{Lucero}},
  \bibinfo{author}{\bibfnamefont{A.~S.} \bibnamefont{Nunes}}, \bibnamefont{and}
  \bibinfo{author}{\bibfnamefont{R.}~\bibnamefont{Sassot}}
  (\bibinfo{year}{2020}), \eprint{2007.08300}.

\bibitem[{\citenamefont{Ravindran et~al.}(2004)\citenamefont{Ravindran, Smith,
  and van Neerven}}]{Ravindran:2003gi}
\bibinfo{author}{\bibfnamefont{V.}~\bibnamefont{Ravindran}},
  \bibinfo{author}{\bibfnamefont{J.}~\bibnamefont{Smith}}, \bibnamefont{and}
  \bibinfo{author}{\bibfnamefont{W.}~\bibnamefont{van Neerven}},
  \bibinfo{journal}{Nucl. Phys. B} \textbf{\bibinfo{volume}{682}},
  \bibinfo{pages}{421} (\bibinfo{year}{2004}), \eprint{hep-ph/0311304}.

\bibitem[{\citenamefont{Zijlstra and van Neerven}(1994)}]{Zijlstra:1993sh}
\bibinfo{author}{\bibfnamefont{E.}~\bibnamefont{Zijlstra}} \bibnamefont{and}
  \bibinfo{author}{\bibfnamefont{W.}~\bibnamefont{van Neerven}},
  \bibinfo{journal}{Nucl. Phys. B} \textbf{\bibinfo{volume}{417}},
  \bibinfo{pages}{61} (\bibinfo{year}{1994}), \bibinfo{note}{[Erratum:
  Nucl.Phys.B 426, 245 (1994), Erratum: Nucl.Phys.B 773, 105--106 (2007),
  Erratum: Nucl.Phys.B 501, 599--599 (1997)]}.

\bibitem[{\citenamefont{Vogt et~al.}(2008)\citenamefont{Vogt, Moch, Rogal, and
  Vermaseren}}]{Vogt:2008yw}
\bibinfo{author}{\bibfnamefont{A.}~\bibnamefont{Vogt}},
  \bibinfo{author}{\bibfnamefont{S.}~\bibnamefont{Moch}},
  \bibinfo{author}{\bibfnamefont{M.}~\bibnamefont{Rogal}}, \bibnamefont{and}
  \bibinfo{author}{\bibfnamefont{J.}~\bibnamefont{Vermaseren}},
  \bibinfo{journal}{Nucl. Phys. B Proc. Suppl.} \textbf{\bibinfo{volume}{183}},
  \bibinfo{pages}{155} (\bibinfo{year}{2008}), \eprint{0807.1238}.

\bibitem[{\citenamefont{Moch et~al.}(2014)\citenamefont{Moch, Vermaseren, and
  Vogt}}]{Moch:2014sna}
\bibinfo{author}{\bibfnamefont{S.}~\bibnamefont{Moch}},
  \bibinfo{author}{\bibfnamefont{J.}~\bibnamefont{Vermaseren}},
  \bibnamefont{and} \bibinfo{author}{\bibfnamefont{A.}~\bibnamefont{Vogt}},
  \bibinfo{journal}{Nucl. Phys. B} \textbf{\bibinfo{volume}{889}},
  \bibinfo{pages}{351} (\bibinfo{year}{2014}), \eprint{1409.5131}.

\bibitem[{\citenamefont{Moch et~al.}(2015)\citenamefont{Moch, Vermaseren, and
  Vogt}}]{Moch:2015usa}
\bibinfo{author}{\bibfnamefont{S.}~\bibnamefont{Moch}},
  \bibinfo{author}{\bibfnamefont{J.}~\bibnamefont{Vermaseren}},
  \bibnamefont{and} \bibinfo{author}{\bibfnamefont{A.}~\bibnamefont{Vogt}},
  \bibinfo{journal}{Phys. Lett. B} \textbf{\bibinfo{volume}{748}},
  \bibinfo{pages}{432} (\bibinfo{year}{2015}), \eprint{1506.04517}.

\bibitem[{\citenamefont{Arratia et~al.}(2020)\citenamefont{Arratia, Furletova,
  Hobbs, Olness, and Sekula}}]{Arratia:2020azl}
\bibinfo{author}{\bibfnamefont{M.}~\bibnamefont{Arratia}},
  \bibinfo{author}{\bibfnamefont{Y.}~\bibnamefont{Furletova}},
  \bibinfo{author}{\bibfnamefont{T.}~\bibnamefont{Hobbs}},
  \bibinfo{author}{\bibfnamefont{F.}~\bibnamefont{Olness}}, \bibnamefont{and}
  \bibinfo{author}{\bibfnamefont{S.~J.} \bibnamefont{Sekula}}
  (\bibinfo{year}{2020}), \eprint{2006.12520}.

\bibitem[{\citenamefont{Kang et~al.}(2020)\citenamefont{Kang, Liu, Mantry, and
  Shao}}]{Kang:2020fka}
\bibinfo{author}{\bibfnamefont{Z.-B.} \bibnamefont{Kang}},
  \bibinfo{author}{\bibfnamefont{X.}~\bibnamefont{Liu}},
  \bibinfo{author}{\bibfnamefont{S.}~\bibnamefont{Mantry}}, \bibnamefont{and}
  \bibinfo{author}{\bibfnamefont{D.~Y.} \bibnamefont{Shao}}
  (\bibinfo{year}{2020}), \eprint{2008.00655}.

\bibitem[{\citenamefont{Catani and Seymour}(1997)}]{Catani:1996vz}
\bibinfo{author}{\bibfnamefont{S.}~\bibnamefont{Catani}} \bibnamefont{and}
  \bibinfo{author}{\bibfnamefont{M.}~\bibnamefont{Seymour}},
  \bibinfo{journal}{Nucl. Phys. B} \textbf{\bibinfo{volume}{485}},
  \bibinfo{pages}{291} (\bibinfo{year}{1997}), \bibinfo{note}{[Erratum:
  Nucl.Phys.B 510, 503--504 (1998)]}, \eprint{hep-ph/9605323}.

\bibitem[{\citenamefont{Borsa et~al.}(2020)\citenamefont{Borsa, de~Florian, and
  Pedron}}]{Borsa:2020ulb}
\bibinfo{author}{\bibfnamefont{I.}~\bibnamefont{Borsa}},
  \bibinfo{author}{\bibfnamefont{D.}~\bibnamefont{de~Florian}},
  \bibnamefont{and} \bibinfo{author}{\bibfnamefont{I.}~\bibnamefont{Pedron}},
  \bibinfo{journal}{Phys. Rev. Lett.} \textbf{\bibinfo{volume}{125}},
  \bibinfo{pages}{082001} (\bibinfo{year}{2020}), \eprint{2005.10705}.

\bibitem[{\citenamefont{Cacciari et~al.}(2015)\citenamefont{Cacciari, Dreyer,
  Karlberg, Salam, and Zanderighi}}]{Cacciari:2015jma}
\bibinfo{author}{\bibfnamefont{M.}~\bibnamefont{Cacciari}},
  \bibinfo{author}{\bibfnamefont{F.~A.} \bibnamefont{Dreyer}},
  \bibinfo{author}{\bibfnamefont{A.}~\bibnamefont{Karlberg}},
  \bibinfo{author}{\bibfnamefont{G.~P.} \bibnamefont{Salam}}, \bibnamefont{and}
  \bibinfo{author}{\bibfnamefont{G.}~\bibnamefont{Zanderighi}},
  \bibinfo{journal}{Phys. Rev. Lett.} \textbf{\bibinfo{volume}{115}},
  \bibinfo{pages}{082002} (\bibinfo{year}{2015}), \bibinfo{note}{[Erratum:
  Phys.Rev.Lett. 120, 139901 (2018)]}, \eprint{1506.02660}.

\bibitem[{\citenamefont{Andreev et~al.}(2015)}]{Andreev:2014wwa}
\bibinfo{author}{\bibfnamefont{V.}~\bibnamefont{Andreev}} \bibnamefont{et~al.}
  (\bibinfo{collaboration}{H1}), \bibinfo{journal}{Eur. Phys. J. C}
  \textbf{\bibinfo{volume}{75}}, \bibinfo{pages}{65} (\bibinfo{year}{2015}),
  \eprint{1406.4709}.

\bibitem[{\citenamefont{Andreev et~al.}(2017)}]{Andreev:2016tgi}
\bibinfo{author}{\bibfnamefont{V.}~\bibnamefont{Andreev}} \bibnamefont{et~al.}
  (\bibinfo{collaboration}{H1}), \bibinfo{journal}{Eur. Phys. J. C}
  \textbf{\bibinfo{volume}{77}}, \bibinfo{pages}{215} (\bibinfo{year}{2017}),
  \eprint{1611.03421}.

\bibitem[{\citenamefont{Abramowicz et~al.}(2010)}]{Abramowicz:2010cka}
\bibinfo{author}{\bibfnamefont{H.}~\bibnamefont{Abramowicz}}
  \bibnamefont{et~al.} (\bibinfo{collaboration}{ZEUS}), \bibinfo{journal}{Eur.
  Phys. J. C} \textbf{\bibinfo{volume}{70}}, \bibinfo{pages}{965}
  (\bibinfo{year}{2010}), \eprint{1010.6167}.

\bibitem[{\citenamefont{Catani and Grazzini}(2007)}]{Catani:2007vq}
\bibinfo{author}{\bibfnamefont{S.}~\bibnamefont{Catani}} \bibnamefont{and}
  \bibinfo{author}{\bibfnamefont{M.}~\bibnamefont{Grazzini}},
  \bibinfo{journal}{Phys. Rev. Lett.} \textbf{\bibinfo{volume}{98}},
  \bibinfo{pages}{222002} (\bibinfo{year}{2007}), \eprint{hep-ph/0703012}.

\bibitem[{\citenamefont{Gehrmann-De~Ridder
  et~al.}(2005)\citenamefont{Gehrmann-De~Ridder, Gehrmann, and
  Glover}}]{GehrmannDeRidder:2005cm}
\bibinfo{author}{\bibfnamefont{A.}~\bibnamefont{Gehrmann-De~Ridder}},
  \bibinfo{author}{\bibfnamefont{T.}~\bibnamefont{Gehrmann}}, \bibnamefont{and}
  \bibinfo{author}{\bibfnamefont{E.}~\bibnamefont{Glover}},
  \bibinfo{journal}{JHEP} \textbf{\bibinfo{volume}{09}}, \bibinfo{pages}{056}
  (\bibinfo{year}{2005}), \eprint{hep-ph/0505111}.

\bibitem[{\citenamefont{Boughezal et~al.}(2012)\citenamefont{Boughezal,
  Melnikov, and Petriello}}]{Boughezal:2011jf}
\bibinfo{author}{\bibfnamefont{R.}~\bibnamefont{Boughezal}},
  \bibinfo{author}{\bibfnamefont{K.}~\bibnamefont{Melnikov}}, \bibnamefont{and}
  \bibinfo{author}{\bibfnamefont{F.}~\bibnamefont{Petriello}},
  \bibinfo{journal}{Phys. Rev. D} \textbf{\bibinfo{volume}{85}},
  \bibinfo{pages}{034025} (\bibinfo{year}{2012}), \eprint{1111.7041}.

\bibitem[{\citenamefont{Czakon}(2010)}]{Czakon:2010td}
\bibinfo{author}{\bibfnamefont{M.}~\bibnamefont{Czakon}},
  \bibinfo{journal}{Phys. Lett. B} \textbf{\bibinfo{volume}{693}},
  \bibinfo{pages}{259} (\bibinfo{year}{2010}), \eprint{1005.0274}.

\bibitem[{\citenamefont{Binoth and Heinrich}(2004)}]{Binoth:2004jv}
\bibinfo{author}{\bibfnamefont{T.}~\bibnamefont{Binoth}} \bibnamefont{and}
  \bibinfo{author}{\bibfnamefont{G.}~\bibnamefont{Heinrich}},
  \bibinfo{journal}{Nucl. Phys. B} \textbf{\bibinfo{volume}{693}},
  \bibinfo{pages}{134} (\bibinfo{year}{2004}), \eprint{hep-ph/0402265}.

\bibitem[{\citenamefont{Anastasiou et~al.}(2004)\citenamefont{Anastasiou,
  Melnikov, and Petriello}}]{Anastasiou:2003gr}
\bibinfo{author}{\bibfnamefont{C.}~\bibnamefont{Anastasiou}},
  \bibinfo{author}{\bibfnamefont{K.}~\bibnamefont{Melnikov}}, \bibnamefont{and}
  \bibinfo{author}{\bibfnamefont{F.}~\bibnamefont{Petriello}},
  \bibinfo{journal}{Phys. Rev. D} \textbf{\bibinfo{volume}{69}},
  \bibinfo{pages}{076010} (\bibinfo{year}{2004}), \eprint{hep-ph/0311311}.

\bibitem[{\citenamefont{Somogyi et~al.}(2007)\citenamefont{Somogyi, Trocsanyi,
  and Del~Duca}}]{Somogyi:2006da}
\bibinfo{author}{\bibfnamefont{G.}~\bibnamefont{Somogyi}},
  \bibinfo{author}{\bibfnamefont{Z.}~\bibnamefont{Trocsanyi}},
  \bibnamefont{and} \bibinfo{author}{\bibfnamefont{V.}~\bibnamefont{Del~Duca}},
  \bibinfo{journal}{JHEP} \textbf{\bibinfo{volume}{01}}, \bibinfo{pages}{070}
  (\bibinfo{year}{2007}), \eprint{hep-ph/0609042}.

\bibitem[{\citenamefont{Stewart et~al.}(2010)\citenamefont{Stewart, Tackmann,
  and Waalewijn}}]{Stewart_2010}
\bibinfo{author}{\bibfnamefont{I.~W.} \bibnamefont{Stewart}},
  \bibinfo{author}{\bibfnamefont{F.~J.} \bibnamefont{Tackmann}},
  \bibnamefont{and} \bibinfo{author}{\bibfnamefont{W.~J.}
  \bibnamefont{Waalewijn}}, \bibinfo{journal}{Physical Review Letters}
  \textbf{\bibinfo{volume}{105}} (\bibinfo{year}{2010}), ISSN
  \bibinfo{issn}{1079-7114},
  \urlprefix\url{http://dx.doi.org/10.1103/PhysRevLett.105.092002}.

\bibitem[{\citenamefont{Antonelli et~al.}(2000)\citenamefont{Antonelli,
  Dasgupta, and Salam}}]{Antonelli:1999kx}
\bibinfo{author}{\bibfnamefont{V.}~\bibnamefont{Antonelli}},
  \bibinfo{author}{\bibfnamefont{M.}~\bibnamefont{Dasgupta}}, \bibnamefont{and}
  \bibinfo{author}{\bibfnamefont{G.~P.} \bibnamefont{Salam}},
  \bibinfo{journal}{JHEP} \textbf{\bibinfo{volume}{02}}, \bibinfo{pages}{001}
  (\bibinfo{year}{2000}), \eprint{hep-ph/9912488}.

\bibitem[{\citenamefont{Dasgupta and Salam}(2002)}]{Dasgupta:2002dc}
\bibinfo{author}{\bibfnamefont{M.}~\bibnamefont{Dasgupta}} \bibnamefont{and}
  \bibinfo{author}{\bibfnamefont{G.~P.} \bibnamefont{Salam}},
  \bibinfo{journal}{JHEP} \textbf{\bibinfo{volume}{08}}, \bibinfo{pages}{032}
  (\bibinfo{year}{2002}), \eprint{hep-ph/0208073}.

\bibitem[{\citenamefont{McCance}(1999)}]{McCance:1999jh}
\bibinfo{author}{\bibfnamefont{G.}~\bibnamefont{McCance}}, in
  \emph{\bibinfo{booktitle}{{Workshop on Monte Carlo Generators for HERA
  Physics (Plenary Starting Meeting)}}} (\bibinfo{year}{1999}), pp.
  \bibinfo{pages}{151--159}, \eprint{hep-ph/9912481}.

\bibitem[{\citenamefont{Nagy and Trocsanyi}(2001)}]{Nagy:2001xb}
\bibinfo{author}{\bibfnamefont{Z.}~\bibnamefont{Nagy}} \bibnamefont{and}
  \bibinfo{author}{\bibfnamefont{Z.}~\bibnamefont{Trocsanyi}},
  \bibinfo{journal}{Phys. Rev. Lett.} \textbf{\bibinfo{volume}{87}},
  \bibinfo{pages}{082001} (\bibinfo{year}{2001}), \eprint{hep-ph/0104315}.

\bibitem[{\citenamefont{{'t Hooft} and Veltman}(1972)}]{THOOFT1972189}
\bibinfo{author}{\bibfnamefont{G.}~\bibnamefont{{'t Hooft}}} \bibnamefont{and}
  \bibinfo{author}{\bibfnamefont{M.}~\bibnamefont{Veltman}},
  \bibinfo{journal}{Nuclear Physics B} \textbf{\bibinfo{volume}{44}},
  \bibinfo{pages}{189 } (\bibinfo{year}{1972}), ISSN \bibinfo{issn}{0550-3213},
  \urlprefix\url{http://www.sciencedirect.com/science/article/pii/0550321372902799}.

\bibitem[{\citenamefont{Breitenlohner and Maison}(1977)}]{Breitenlohner:1977hr}
\bibinfo{author}{\bibfnamefont{P.}~\bibnamefont{Breitenlohner}}
  \bibnamefont{and} \bibinfo{author}{\bibfnamefont{D.}~\bibnamefont{Maison}},
  \bibinfo{journal}{Commun. Math. Phys.} \textbf{\bibinfo{volume}{52}},
  \bibinfo{pages}{11} (\bibinfo{year}{1977}).

\bibitem[{\citenamefont{Vogelsang}(1996)}]{Vogelsang:1996im}
\bibinfo{author}{\bibfnamefont{W.}~\bibnamefont{Vogelsang}},
  \bibinfo{journal}{Nucl. Phys. B} \textbf{\bibinfo{volume}{475}},
  \bibinfo{pages}{47} (\bibinfo{year}{1996}), \eprint{hep-ph/9603366}.

\bibitem[{\citenamefont{Vermaseren et~al.}(2005)\citenamefont{Vermaseren, Vogt,
  and Moch}}]{Vermaseren:2005qc}
\bibinfo{author}{\bibfnamefont{J.}~\bibnamefont{Vermaseren}},
  \bibinfo{author}{\bibfnamefont{A.}~\bibnamefont{Vogt}}, \bibnamefont{and}
  \bibinfo{author}{\bibfnamefont{S.}~\bibnamefont{Moch}},
  \bibinfo{journal}{Nucl. Phys. B} \textbf{\bibinfo{volume}{724}},
  \bibinfo{pages}{3} (\bibinfo{year}{2005}), \eprint{hep-ph/0504242}.

\bibitem[{\citenamefont{Zijlstra and van Neerven}(1992)}]{Zijlstra:1992qd}
\bibinfo{author}{\bibfnamefont{E.}~\bibnamefont{Zijlstra}} \bibnamefont{and}
  \bibinfo{author}{\bibfnamefont{W.}~\bibnamefont{van Neerven}},
  \bibinfo{journal}{Nucl. Phys. B} \textbf{\bibinfo{volume}{383}},
  \bibinfo{pages}{525} (\bibinfo{year}{1992}).

\bibitem[{\citenamefont{Butterworth et~al.}(2016)}]{Butterworth:2015oua}
\bibinfo{author}{\bibfnamefont{J.}~\bibnamefont{Butterworth}}
  \bibnamefont{et~al.}, \bibinfo{journal}{J. Phys. G}
  \textbf{\bibinfo{volume}{43}}, \bibinfo{pages}{023001}
  (\bibinfo{year}{2016}), \eprint{1510.03865}.

\bibitem[{\citenamefont{de~Florian et~al.}(2019)\citenamefont{de~Florian,
  Lucero, Sassot, Stratmann, and Vogelsang}}]{deFlorian:2019zkl}
\bibinfo{author}{\bibfnamefont{D.}~\bibnamefont{de~Florian}},
  \bibinfo{author}{\bibfnamefont{G.~A.} \bibnamefont{Lucero}},
  \bibinfo{author}{\bibfnamefont{R.}~\bibnamefont{Sassot}},
  \bibinfo{author}{\bibfnamefont{M.}~\bibnamefont{Stratmann}},
  \bibnamefont{and}
  \bibinfo{author}{\bibfnamefont{W.}~\bibnamefont{Vogelsang}},
  \bibinfo{journal}{Phys. Rev. D} \textbf{\bibinfo{volume}{100}},
  \bibinfo{pages}{114027} (\bibinfo{year}{2019}), \eprint{1902.10548}.

\bibitem[{\citenamefont{Currie et~al.}(2017)\citenamefont{Currie, Gehrmann,
  Huss, and Niehues}}]{Currie:2017tpe}
\bibinfo{author}{\bibfnamefont{J.}~\bibnamefont{Currie}},
  \bibinfo{author}{\bibfnamefont{T.}~\bibnamefont{Gehrmann}},
  \bibinfo{author}{\bibfnamefont{A.}~\bibnamefont{Huss}}, \bibnamefont{and}
  \bibinfo{author}{\bibfnamefont{J.}~\bibnamefont{Niehues}},
  \bibinfo{journal}{JHEP} \textbf{\bibinfo{volume}{07}}, \bibinfo{pages}{018}
  (\bibinfo{year}{2017}), \eprint{1703.05977}.

\bibitem[{\citenamefont{Catani and Webber}(1997)}]{Catani:1997xc}
\bibinfo{author}{\bibfnamefont{S.}~\bibnamefont{Catani}} \bibnamefont{and}
  \bibinfo{author}{\bibfnamefont{B.}~\bibnamefont{Webber}},
  \bibinfo{journal}{JHEP} \textbf{\bibinfo{volume}{10}}, \bibinfo{pages}{005}
  (\bibinfo{year}{1997}), \eprint{hep-ph/9710333}.

\bibitem[{\citenamefont{Frixione and Ridolfi}(1997)}]{Frixione:1997ks}
\bibinfo{author}{\bibfnamefont{S.}~\bibnamefont{Frixione}} \bibnamefont{and}
  \bibinfo{author}{\bibfnamefont{G.}~\bibnamefont{Ridolfi}},
  \bibinfo{journal}{Nucl. Phys. B} \textbf{\bibinfo{volume}{507}},
  \bibinfo{pages}{315} (\bibinfo{year}{1997}), \eprint{hep-ph/9707345}.

\bibitem[{\citenamefont{Graudenz}(1997)}]{Graudenz:1997gv}
\bibinfo{author}{\bibfnamefont{D.}~\bibnamefont{Graudenz}}
  (\bibinfo{year}{1997}), \eprint{hep-ph/9710244}.

\bibitem[{\citenamefont{Dasgupta and Salam}(2001)}]{Dasgupta:2001sh}
\bibinfo{author}{\bibfnamefont{M.}~\bibnamefont{Dasgupta}} \bibnamefont{and}
  \bibinfo{author}{\bibfnamefont{G.}~\bibnamefont{Salam}},
  \bibinfo{journal}{Phys. Lett. B} \textbf{\bibinfo{volume}{512}},
  \bibinfo{pages}{323} (\bibinfo{year}{2001}), \eprint{hep-ph/0104277}.

\bibitem[{\citenamefont{Frixione et~al.}(1996)\citenamefont{Frixione, Kunszt,
  and Signer}}]{Frixione:1995ms}
\bibinfo{author}{\bibfnamefont{S.}~\bibnamefont{Frixione}},
  \bibinfo{author}{\bibfnamefont{Z.}~\bibnamefont{Kunszt}}, \bibnamefont{and}
  \bibinfo{author}{\bibfnamefont{A.}~\bibnamefont{Signer}},
  \bibinfo{journal}{Nucl. Phys. B} \textbf{\bibinfo{volume}{467}},
  \bibinfo{pages}{399} (\bibinfo{year}{1996}), \eprint{hep-ph/9512328}.

\bibitem[{\citenamefont{Mangano and Parke}(1991)}]{Mangano:1990by}
\bibinfo{author}{\bibfnamefont{M.~L.} \bibnamefont{Mangano}} \bibnamefont{and}
  \bibinfo{author}{\bibfnamefont{S.~J.} \bibnamefont{Parke}},
  \bibinfo{journal}{Phys. Rept.} \textbf{\bibinfo{volume}{200}},
  \bibinfo{pages}{301} (\bibinfo{year}{1991}), \eprint{hep-th/0509223}.

\end{thebibliography}
\clearpage

\appendix

\section{Dipole bug in {\tt DISENT}}
\label{sec_apendixbug}

The presence of a bug in the gluon channel in {\tt DISENT} was reported long ago in \cite{Antonelli:1999kx,Dasgupta:2002dc, McCance:1999jh,Nagy:2001xb}, particularly while studying the  event shape distributions in DIS. After a careful analysis, along with an extensive comparison with {\tt DISASTER} \cite{Graudenz:1997gv} (a code which showed good agreement with resummed event shape calculations), and also by writing independent codes, we found that the Born matrix element used in one of the dipole subtraction terms in the gluon channel had the momentum of two final-state partons interchanged, leading to the reported discrepancies. Due to the nature of the bug, it turns out to produce noticeable differences only in certain extreme regions of the phase space, and remains within the typical statistical uncertainties of the calculations in many others.

\begin{figure}[h!]
\begin{center}
\includegraphics [width=0.75\textwidth]{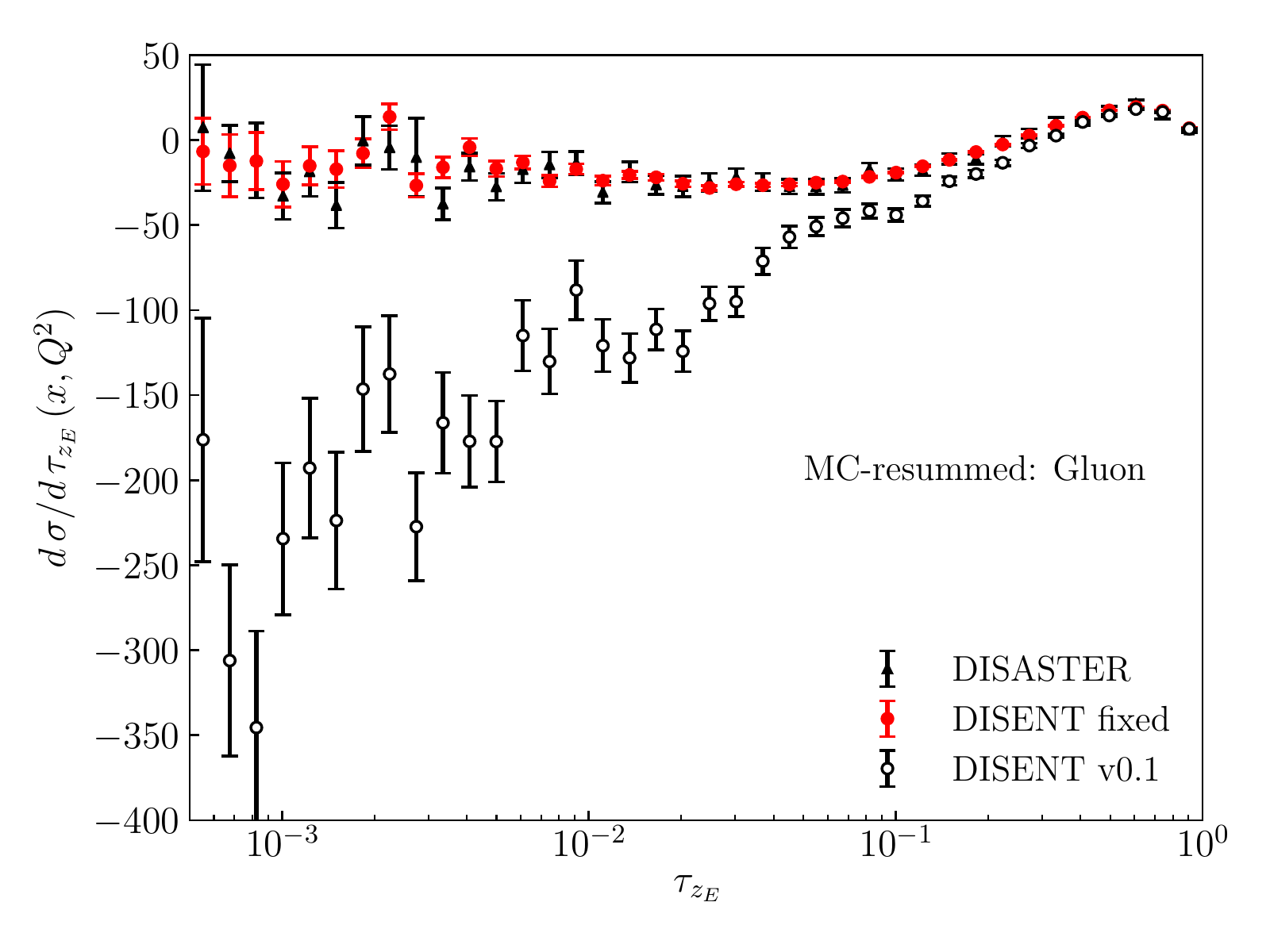} 
\end{center}
 \caption{Difference between the $\mathcal{O}(\alpha_{S}^{2})$ coefficient for the fixed-order Monte Carlo calculation and the expansion of the resummed calculation for the gluonic contribution to $\tau_{z_{E}}$, using {\tt DISASTER}, the original version of {\tt DISENT} and its fixed version.\label{fig} }
\end{figure}

We have checked that the fixed counterterm actually corrects the reported disagreement between {\tt DISENT} and {\tt DISASTER} in the event shapes, as well as the differences between {\tt DISENT} and the analytical calculation for logarithmically enhanced terms. As an example, we present in Fig.~\ref{fig} the difference between the $\mathcal{O}(\alpha_{S}^{2})$ coefficient for the fixed-order Monte Carlo calculation and the expansion of the resummed calculation for the gluonic contribution to $\tau_{z_{E}}$, using {\tt DISASTER}, the v$0.1$ version of {\tt DISENT} and its fixed version. The event shape presented was calculated (at $x_{bj}=0.0039$, $Q^2=7.5$ GeV and $y=0.001$) with the programs {\tt Dispatch} and {\tt DISresum}, written by Salam et al \cite{Antonelli:1999kx,Dasgupta:2002dc,Dasgupta:2001sh}. Similar results are obtained in the case of $\tau_{z_{Q}}$. We also found agreement between {\tt DISASTER} and the modified version of {\tt DISENT} for the quark channel, and for other event shapes.

\section{Spin correlations}
\label{sec_apendixcorr}

As it was pointed out in section~\ref{sec_subtraction}, while, in principle, the initial-state dipole factorization formula for an $n$-parton scattering involves \textit{spin correlations} between spin-dependent kernels and $(n-1)$-particle matrix elements, those correlations cancel in polarized cross sections. The appearance of such correlations, as well as their cancellation in the polarized case, are simpler to deduce within the helicity amplitudes formalism. Since the calculation of polarized cross sections involve differences between the helicity states of incoming particles, we consider an $n$-particle scattering of the form $$p_{a}^{\lambda_{a}}+p_{1}^{\lambda_{1}}\rightarrow p_{2}^{\lambda_2}+X_{f}^{\{ \lambda_{X_{f}}\}},$$
with $p_{1}$ and $p_{2}$ representing an incoming and an outgoing parton, respectively. The superscript $\lambda_{i}$ is used to indicate the helicity state of the $i$-th particle. $p_{a}$ represents an additional incoming particle, while $X_{f}$ is used for the remaining $n-3$ particles involved in the scattering. In the collinear limit between the partons 1 and 2, and following the notation from \cite{Frixione:1995ms}, the $n$-particle amplitude satisfies the strict factorization formula

\begin{equation}\label{eq:fact_formula}
    \mathcal{M}^{n}(\lambda_{1},\lambda_{2},\lambda_{a},\{\lambda_{X_{f}}\})\xrightarrow{1\parallel 2}\,g_{s}\sum_{c_{e}}\sum_{\lambda_{e}} \mathcal{C}(c_{e},c_{1},c_{2})\, S_{1e} ^{\lambda_{1},\lambda_{2},\lambda_{e}}(z)\,\times\, \mathcal{M}^{n-1}(\lambda_{e},\lambda_{a},\{\lambda_{X_{f}}\}),
\end{equation}

\noindent where $S_{1e}^{\lambda_{1},\lambda_{2},\lambda_{e}}$ represents the splitting function with fixed helicities for the process $p_{1}\rightarrow p_{2}+p_{e}$, and $\mathcal{C}(c_{e},c_{1},c_{2})$ is the associated color structure. Schematically, the collinear limit can be represented as

\begin{figure}[h]
 \epsfig{figure= 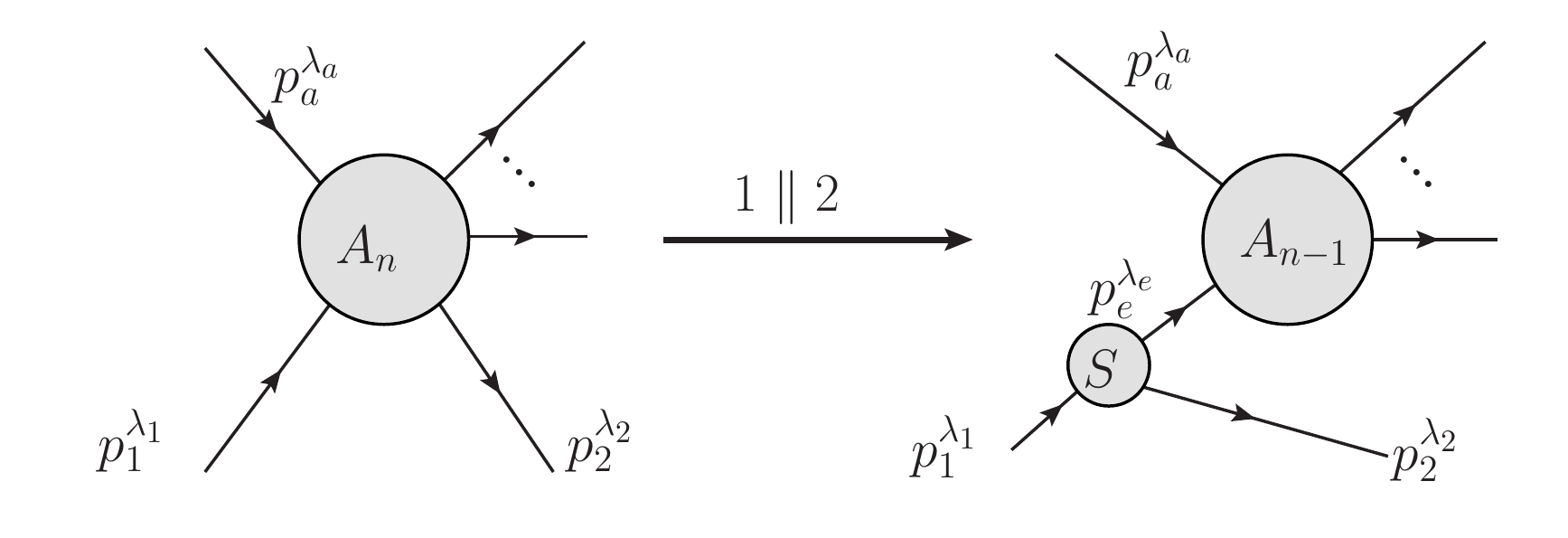 ,width=0.75\textwidth}
  \caption{Collinear behavior for the n-particle amplitude $\mathcal{M}^{n}$ in the limit $1\parallel 2$}
\end{figure}

It should be noted that, even after fixing all the external particles helicities, the factorization at the \textit{amplitude} level involves the summation over the helicity states $\lambda_{e}$ of the intermediate parton. The case in which $p_{e}$ is a quark is trivial, since helicity conservation at the vertex implies that one of the terms in the sum over $\lambda_{e}$ is zero. The case with an intermediate gluon is, however, more involved. The exact factorization is lost at the squared-amplitude level, through the appearance of interference terms between the different helicities in the propagator. Those interference terms give rise to, precisely, the spin correlations noted in the dipole factorization formula. The exact form of the correlation terms can be easily obtained by squaring Eq.~(\ref{eq:fact_formula}):

\begin{equation}\label{eq:correlations}
    |\mathcal{M}^{n}(\lambda_{1},\lambda_{2},\lambda_{a},\{\lambda_{X_{f}}\})|^{2}\xrightarrow{1\parallel 2}\,|\mathcal{N}^{n}(\lambda_{1},\lambda_{2},\lambda_{a},\{\lambda_{X_{f}}\})|^{2}+\mathcal{R}(\lambda_{1},\lambda_{2},\lambda_{a},\{\lambda_{X_{f}}\}),
\end{equation}

\noindent where

\begin{equation}
\begin{split}
    |\mathcal{N}^{n}(\lambda_{1},\lambda_{2},\lambda_{a},\{\lambda_{X_{f}}\})|^{2} & =  g_{S}^{2}\,C\,|S_{1g}^{\lambda_{1}\lambda_{2}+}(z)|^{2}\,|\mathcal{M}^{n-1}(+,\lambda_{a},\{\lambda_{X_{f}}\})|^{2}\\
    &+ g_{S}^{2}\,C\,|S_{1g}^{\lambda_{1}\lambda_{2}-}(z)|^{2}\,|\mathcal{M}^{n-1}(-,\lambda_{a},\{\lambda_{X_{f}}\})|^{2},
\label{eq:corr_css}
\end{split}
\end{equation}

\noindent and the interference term is given by

\begin{equation}\label{eq:correlation_term}
\begin{split}
\mathcal{R}(\lambda_{1},\lambda_{2},\lambda_{a},\{\lambda_{X_{f}}\})& =  2 g_{S}^{2}\,C\,\operatorname{Re}\bigg\{S_{1g}^{\lambda_{1}\lambda_{2}+}(z)\big(S_{1g}^{\lambda_{1}\lambda_{2}-}(z)\big)^{*}\\
& \times \mathcal{M}^{n-1}(+,\lambda_{a},\{\lambda_{X_{f}}\})\big(\mathcal{M}^{n-1}(-,\lambda_{a},\{\lambda_{X_{f}}\})\big)^{*}\bigg\}.
\end{split}
\end{equation}

\noindent In Eqs.~(\ref{eq:corr_css}) and (\ref{eq:correlation_term}) we introduced the short-hand notation for the color factor

\begin{equation}
    C=\sum_{c_{1},c_{2},c_{e},c_{e'}}\frac{1}{N_{c_{1}}}\,\mathcal{C}(c_{e},c_{1},c_{2})\,\mathcal{C}^{*}(c_{e'},c_{1},c_{2}),
\end{equation}

 \noindent with $1/N_{c_{1}}$ denoting the average over the initial parton colors. For the relevant cases, and using the normalization from \cite{Mangano:1990by}, $C$ can take the values $2 C_{A}$ and $C_{F}$, for an initial gluon and quark, respectively. Notice that the interference term depends on the initial parton helicity $\lambda_{1}$ only through the spin-dependent kernels $S_{1g}^{\lambda_{1}\lambda_{2}\lambda_{e}}$. In the calculation of the unpolarized (polarized) cross section, we can then write:

\begin{equation}
\begin{split}
    (\Delta)\sigma=\sum_{\lambda_{2},\lambda_{a},\{\lambda_{X_{f}}\}} \frac{(\lambda_{a})}{4}\times\Big[& |\mathcal{M}^{n}(+,\lambda_{2},\lambda_{a},\{\lambda_{X_{f}}\})|^{2}+(-)|\mathcal{M}^{n}(-,\lambda_{2},\lambda_{a},\{\lambda_{X_{f}}\})|^{2}\Big],
\end{split}
\end{equation}

\noindent where the helicity factor $(\lambda_{a})$ should only be considered in the polarized case. Using Eq.~(\ref{eq:correlations}), the unpolarized (polarized) cross section can in turn be expressed as

\begin{equation}\label{eq:sigmapol_completo}
\begin{split}
    (\Delta)\sigma=\sum_{\lambda_{2},\lambda_{a},\{\lambda_{X_{f}}\}}& \frac{(\lambda_{a})}{4}\times\Big[ |\mathcal{N}^{n}(+,\lambda_{2},\lambda_{a},\{\lambda_{X_{f}}\})|^{2}+(-)|\mathcal{N}^{n}(-,\lambda_{2},\lambda_{a},\{\lambda_{X_{f}}\})|^{2}\\
    &+\mathcal{R}(+,\lambda_{2},\lambda_{a},\{\lambda_{X_{f}}\})+(-)\mathcal{R}(-,\lambda_{2},\lambda_{a},\{\lambda_{X_{f}}\})\Big].
\end{split}
\end{equation}

We can then write explicitly the correlation terms in the second line of Eq.~(\ref{eq:sigmapol_completo}) and sum over the polarizations of the final-state particles $X_f$

\begin{equation}\label{eq:cancellations}
\begin{split}
    \sum_{\lambda_{2},\{\lambda_{X_{f}}\}}\Big[\mathcal{R}(+,\lambda_{2},\lambda_{a},& \{\lambda_{X_{f}}\})+(-)\mathcal{R}(-,\lambda_{2},\lambda_{a},\{\lambda_{X_{f}}\})\Big]\\
    &=2 g_{S}^{2}\,C\,\operatorname{Re}\bigg\{S_{1g}^{+-+}(z)\big(S_{1g}^{+--}(z)\big)^{*}
\times \mathcal{M}^{n-1}(+,\lambda_{a})\big(\mathcal{M}^{n-1}(-,\lambda_{a})\big)^{*}\bigg\}\\
&+2 g_{S}^{2}\,C\,\operatorname{Re}\bigg\{S_{1g}^{+++}(z)\big(S_{1g}^{++-}(z)\big)^{*}
\times \mathcal{M}^{n-1}(+,\lambda_{a})\big(\mathcal{M}^{n-1}(-,\lambda_{a})\big)^{*}\bigg\}\\
&+(-)2 g_{S}^{2}\,C\,\operatorname{Re}\bigg\{S_{1g}^{-++}(z)\big(S_{1g}^{-+-}(z)\big)^{*}
\times \mathcal{M}^{n-1}(+,\lambda_{a})\big(\mathcal{M}^{n-1}(-,\lambda_{a})\big)^{*}\bigg\}\\
&+(-)2 g_{S}^{2}\,C\,\operatorname{Re}\bigg\{S_{1g}^{--+}(z)\big(S_{1g}^{---}(z)\big)^{*}
\times \mathcal{M}^{n-1}(+,\lambda_{a})\big(\mathcal{M}^{n-1}(-,\lambda_{a})\big)^{*}\bigg\}\\
&=\begin{cases}
4 g_{S}^{2}\,C\,\operatorname{Re}\bigg\{ \bigg(S_{1g}^{+-+}(z)\big(S_{1g}^{+--}(z)\big)^{*}
+S_{1g}^{+++}(z)\big(S_{1g}^{++-}(z)\big)^{*}\bigg)\\
\qquad\qquad\qquad\times \mathcal{M}^{n-1}(+,\lambda_{a})\big(\mathcal{M}^{n-1}(-,\lambda_{a})\big)^{*}\bigg\} & \mbox{unpolarized}\\
0 & \mbox{polarized}
\end{cases}.
\end{split}
\end{equation}

\noindent In the last step we used that parity conservation implies that

\begin{equation}
\begin{split}
&S_{1g}^{++-}(z)=\big(S_{1g}^{--+}(z)\big)^{*},\qquad S_{1g}^{+-+}(z)=\big(S_{1g}^{-+-}(z)\big)^{*}, \\
&S_{1g}^{+++}(z)=\big(S_{1g}^{---}(z)\big)^{*},\qquad S_{1g}^{+--}(z)=\big(S_{1g}^{-++}(z)\big)^{*},\\
\end{split}
\end{equation}

\noindent so all the interference terms in Eq.(~\ref{eq:cancellations}) cancel each other in the polarized cross section. Thus, in the polarized case we simply obtain

\begin{equation}
\begin{split}
\Delta\sigma &=\sum_{\lambda_{2},\lambda_{a},\{\lambda_{X_{f}}\}}\frac{\lambda_{a}}{4}\Big[ |\mathcal{N}^{n}(+,\lambda_{2},\lambda_{a},\{\lambda_{X_{f}}\})|^{2}-|\mathcal{N}^{n}(-,\lambda_{2},\lambda_{a},\{\lambda_{X_{f}}\})|^{2}\Big]\\
    &=\sum_{\lambda_{2},\lambda_{a},\{\lambda_{X_{f}}\}}\frac{\lambda_{a}}{4}\Big[ g_{S}^{2}\,C\,|S_{1g}^{+\lambda_{2}+}(z)|^{2}\,|\mathcal{M}^{n-1}(+,\lambda_{a},\{\lambda_{X_{f}}\})|^{2}\\
     &\qquad\qquad\qquad+ g_{S}^{2}\,C\,|S_{1g}^{+\lambda_{2}-}(z)|^{2}\,|\mathcal{M}^{n-1}(-,\lambda_{a},\{\lambda_{X_{f}}\})|^{2}\\
    &\qquad\qquad\qquad-g_{S}^{2}\,C\,|S_{1g}^{-\lambda_{2}+}(z)|^{2}\,|\mathcal{M}^{n-1}(+,\lambda_{a},\{\lambda_{X_{f}}\})|^{2}\\
    &\qquad\qquad\qquad- g_{S}^{2}\,C\,|S_{1g}^{-\lambda_{2}-}(z)|^{2}\,|\mathcal{M}^{n-1}(-,\lambda_{a},\{\lambda_{X_{f}}\})|^{2}\Big]\\
    &=\sum_{\{\lambda_{X_{f}}\}}\,\frac{4\pi \alpha_{S}^{2}}{p_{1}\cdot p_{2}}\,\Delta P^<_{1g}(z)\,|\Delta\mathcal{M}^{n-1}(\{\lambda_{X_{f}}\})|^{2}=\frac{4\pi \alpha_{S}^{2}}{p_{1}\cdot p_{2}}\,\Delta P^<_{1g}(z)\,|\Delta\mathcal{M}^{n-1}|^{2},
\end{split}
\end{equation}

\noindent where we have used that the polarized Altarelli-Parisi kernels for $z<1$, $\Delta P^<_{1j}(z)$, can be obtained from the helicity-dependent kernels as
\begin{equation}
\frac{1}{2\,p_{i}\cdot p_{j}}\,\Delta P^{<}_{ij}(z)=\sum_{\lambda_{i},\lambda_{j}} \lambda_{i}\lambda_{j}\,|S_{ij}^{\lambda_{i}+\lambda_{j}}(z)|^{2}=\sum_{\lambda_{i},\lambda_{j}} \lambda_{i}\lambda_{j}\,|S_{ij}^{\lambda_{i}-\lambda_{j}}(z)|^{2},
\end{equation}

\noindent and defined 

\begin{equation}
|\Delta\mathcal{M}^{n-1}|^{2}\equiv \sum_{\lambda_{e},\lambda_{a},\{\lambda_{X_{f}}\}}\,\frac{\lambda_{e}\,\lambda_{a}}{4}|\mathcal{M}^{n-1}(\lambda_{e},\lambda_{a},\{\lambda_{X_{f}}\})|^{2}.
\end{equation}

For the unpolarized cross section, a similar procedure leads to \cite{Frixione:1995ms}:

\begin{equation}
\sigma=\frac{4\pi \alpha_{S}^{2}}{p_{1}\cdot p_{2}}\, P^<_{1g}(z)\,|\mathcal{M}^{n-1}|^{2}+\frac{4\pi \alpha_{S}^{2}}{p_{1}\cdot p_{2}}\, Q^<_{1g}(z)\,|\widetilde{\mathcal{M}}^{n-1}|^{2}, 
\end{equation}

\noindent where in the second term, the one that originates from spin correlations, we defined

\begin{equation}
|\widetilde{\mathcal{M}}^{n-1}|^{2}= \operatorname{Re}\bigg\{ \frac{\langle12\rangle}{[12]}\times \sum_{\lambda_{a}}\frac{1}{2}\mathcal{M}^{n-1}(+,\lambda_{a})\big(\mathcal{M}^{n-1}(-,\lambda_{a})\big)^{*}\bigg\},   
\end{equation}

\noindent while the factor $Q^<_{1g}(z)$ takes the values $-4C_{A}z(1-z)$ and $4T_{R}z(1-z)$ for an initial gluon and quark, respectively.

\end{document}